\documentclass{aa}  
\usepackage{natbib}
\bibpunct{(}{)}{;}{a}{}{,} 
\usepackage{graphicx}
\usepackage{txfonts}
\usepackage[colorlinks=true, citecolor=blue, urlcolor=blue]{hyperref}
\usepackage{placeins}
\usepackage{color, soul}

\begin{document}

   \title{LHAASO J2226+6057 as a pulsar wind nebula}
    

   \author{Agnibha De Sarkar\inst{1,2},
          Wei Zhang\inst{2,4},
          Jonatan Mart\'in\inst{3},
          Diego F. Torres\inst{2,4,5},
          Jian Li \inst{6,7},
          \&
          Xian Hou \inst{8,9}
                 }

   \institute{Raman Research Institute, Bengaluru 560080, Karnataka, India
        \and      
             Institute of Space Sciences (ICE, CSIC), Campus UAB, Carrer de Magrans s/n, E-08193 Barcelona, Spain
        \and  
         INAF - Osservatorio Astrofisico di Arcetri, Largo E. Fermi 5, I-50125 Firenze, Italy 
         \and
            Institut d’Estudis Espacials de Catalunya (IEEC), E-08034 Barcelona, Spain
        \and    
            Instituci\'o Catalana de Recerca i Estudis Avan\k{c}ats (ICREA), Barcelona, Spain
        \and 
        CAS Key Laboratory for Research in Galaxies and Cosmology, Department of Astronomy, University of Science and Technology of China, Hefei, China 
        \and 
School of Astronomy and Space Science, University of Science and Technology of China, Hefei, China
\and 
Yunnan Observatories, Chinese Academy of Sciences, Kunming 650216, China
\and 
Key Laboratory for the Structure and Evolution of Celestial Objects, Chinese Academy of Sciences, Kunming 650216, China\\ 
             \email{agnibha@rri.res.in}
             }

   \date{Received YYYY; accepted ZZZZ}

 
  \abstract
   {The Large High Altitude Air Shower Observatory has reported the detection of cosmic-ray sources in Milky Way that can accelerate particles up to PeV (= 10$^{15}$ eV) energies. These sources, so called ``PeVatrons'', are mostly unidentified. Several classes of sources, such as supernova remnants, pulsar wind nebula, or young stellar clusters can potentially be the counterparts of these PeVatrons. 
   }
   {The aim of this work is to study a pulsar wind nebula interpretation of one of these PeVatrons, LHAASO J2226+6057, which has a relatively well covered multi-frequency spectrum.}
   {We have performed a leptonic, time-dependent modeling of the pulsar wind nebula (PWN) associated with PSR J2229+6114 considering a time-energy-dependent diffusion-loss equation. Injection, energy losses, as well as escape of particles were considered to balance the time-dependent lepton population. We have also included the dynamics of the PWN and the associated supernova remnant (SNR) and their interaction via the reverse shock to study the reverberation phase of the system.}
   {
   We have considered different values of braking index ($n$) and true age ($t_{age}$) for the fitting of the multi-wavelength (MWL) spectral energy distribution (SED) of LHAASO J2226+6057. The best-fit PWN model parameters and their 1$\sigma$ confidence intervals were evaluated. We have also demonstrated the impact of reverberation on the MWL SED with increasing time. Additionally, we have discussed the resultant large radius and low magnetic field associated with the PWN in question, as caveats for the possible physical connection of the pulsar as the origin of this high energy source.}
   {}

   \titlerunning{LHAASO J2226+6057}
   \authorrunning{De Sarkar et al.}
   \maketitle
%

\section{Introduction}

Recent observations by state-of-the-art observatories such as the
Large High Altitude Air Shower Observatory (LHAASO), Tibet AS$\gamma$, the 
High Altitude Water Cherenkov (HAWC), and others, have paved the way for the detection of multiple Galactic ultra-high energy (UHE; E$_\gamma \ge$ 100 TeV) gamma ray sources \citep{abeysekara20, amenomori19, amenomori21, cao21}. 
%
%
%
Upcoming observatories such as the Cherenkov Telescope Array \citep[CTA;][]{CTA19} and the Southern Wide-field Gamma-ray Observatory \citep[SWGO;][]{SWGO19} will be of importance to identify and characterize these PeVatrons: Galactic CR sources that accelerate particles up to PeV energies.

LHAASO is a state-of-the-art dual-task facility designed for CRs and gamma ray studies at few hundred GeV to few PeV, located at 4410 m above sea level in China \citep{cao10}. The recent data reported by the LHAASO observatory shows the existence of 12 significantly detected sources ($>$ 7$\sigma$) that emit gamma rays with energies above several hundred TeVs \citep{cao21}. 
 Most of the sources reported by LHAASO have significantly extended gamma ray emission 
regions up to $\sim$ 1$^\circ$. The very high energy (VHE; 100 GeV $\le$ E$_\gamma$ $\le$ 100 TeV) counterparts of these sources residing in the Galactic plane have been associated with pulsar wind nebulae (PWNe), based on the spatial proximity with highly energetic pulsars and typically extended morphological features \citep{hess18}. 
It has already been posited that UHE gamma ray emission spatially coincident or in very close proximity of energetic pulsars with high spin-down luminosity ($\Dot{E} >$ 10$^{36}$ erg s$^{-1}$) may be a universal feature, see e.g., \cite{albert21}. Moreover, the Crab nebula, associated with pulsar PSR B0531+21, was confirmed to be a PeVatron by recent LHAASO observations \citep{cao21}. Bearing all of this in mind, it is natural to consider PWNe as  possible Galactic PeVatrons, from which UHE gamma rays are detected.

PWNe, considered to be one of the most efficient lepton accelerators in Galaxy, are powered by highly energetic pulsars. 
Pulsars dissipate most of their rotational energy via the injection of ultrarelativistic electron-positron pairs, which form a cold, ultrarelativistic wind of particles. Since the bulk velocity of this ultrarelativistic wind is supersonic with respect to the ambient medium, this wind creates a termination shock. Injected particles can be accelerated to very high energies at this termination shock. 
The accelerated leptons can then interact with the ambient matter, photon fields, and the magnetic field through Bremsstrahlung, inverse-Compton (IC) and synchrotron processes. The cooling of the accelerated leptons results in a MWL spectrum ranging from radio to gamma ray energies. 
Here, we take LHAASO J2226+6057 as our source of interest, as this is the only UHE gamma ray source for which data have been observed across radio, X-ray, and gamma ray energy ranges.  
It has been observed that LHAASO J2226+6057 is situated in a complex morphological region. Due to the close spatial proximity of Boomerang PWN, as well as SNR G106.3+2.7 and associated molecular clouds (MCs), it is hard to confirm the exact source responsible for UHE gamma ray emission observed. From the observations of Imaging Air Cherenkov Telescopes (IACTs) such as MAGIC and VERITAS, it was found that the emission region was divided into two morphological regions; the head and the tail. 
The head region contains Boomerang PWN and PSR J2229+6114, and the tail region contains VER J2227+608, which is likely to be associated with SNR G106.3+2.7 and the MCs in the region. Faint and diffuse radio and X-ray emission were also observed from the tail region. It is possible that the emission from Boomerang PWN illuminates the head region, whereas hadronic interaction occurring between the SNR and the MCs are responsible for the gamma ray emission in the tail. 
Although it is difficult to confirm which source is actually responsible for the UHE gamma ray emission, comprehensive studies are needed to explore both scenarios. New data from upcoming observations with high angular resolution by MAGIC and VERITAS will be crucial to shed new light on this source.
We discuss the features of LHAASO J2226+6057, as well as the previous works done for this source in Sec. \ref{J2226+6057}. Then we comment on the PWN model used in this work in Sec. \ref{sec2}, and present the results in Sec. \ref{sec3}. We finally conclude in Sec. \ref{sec5}.

\section{LHAASO J2226+6057 features}\label{J2226+6057}

LHAASO J2226+6057 was detected at RA = 336.75$^\circ$ and Decl. = 60.95$^\circ$ with a significance of 13.6$\sigma$ above 100 TeV. 
Its gamma ray spectrum reaches up to a maximum energy of 0.57 $\pm$ 0.19 PeV \citep{cao21}. This source is spatially associated with the supernova remnant SNR G106.3+2.7, as well as the pulsar J2229+6114 and its wind nebula, known as the ``Boomerang'' nebula \citep{kothes01}. 
PSR J2229+6114 is a bright gamma ray pulsar with a spin period of 51.6 ms, a characteristic age of 10460 yr and a spin-down luminosity of 2.2 $\times$ 10$^{37}$ erg s$^{-1}$ \citep{halpern01}. 
Pulsed GeV gamma ray emission from this pulsar was detected by \textit{Fermi}-LAT \citep{abdo09}. 
In the VHE range, SNR G106.3+2.7 was observed by VERITAS as VER J2227+608, located 0.4$^{\circ}$ away from PSR J2229+6114 \citep{acciari09}. GeV gamma rays \citep{xin19}, diffuse non-thermal X-rays \citep{fujita21}, as well as radio data \citep{pineault2000} have also been observed from the source region. 
The distance of the source suffers from great uncertainty. It was estimated to be 7.5 kpc \citep[based on pulsar dispersion measure;][]{abdo09}, 3 kpc \citep[based on X-ray absorption measurements;][]{halpern01}, and 0.8 kpc \citep[based on measurements of radial velocities of atomic hydrogen and molecular material;][]{kothes01}. In this work we consider the distance to be 3 kpc, similar to \cite{joshi22} and \cite{yu22}.


The possible connection between LHAASO J2226+6057 and PSR J2229+6114 has previously been studied by \cite{breuhaus22}, \cite{joshi22} and \cite{yu22}. 

\cite{breuhaus22} has performed a time-independent one-zone treatment, a steady-state leptonic scenario from the PWN, to explain only the highest energy gamma ray data. The evolution of the source was neglected, as well as MWL data were not used. 
\cite{joshi22} have used a time-dependent one-zone leptonic scenario from the PWN. However, the authors did not consider the effect of escape for LHAASO J2226+6057, which they accounted only for LHAASO J1908+0621, nor explored the effect of age and braking index on the evolution of the injected leptonic population. The impact of SNR reverse shock and its effects on PWN radius evolution were not taken into account, assuming that such effects are important only if the age of the PWN is greater than 10 kyr, which is not necessarily the case, see, e.g., \cite{martin16}. 
\cite{yu22} have also performed a similar study. The authors had argued that a distorted nebula, created due to the impact of SNR reverse shock, is responsible for the GeV gamma ray emission observed from the source region by \cite{xin19}. Their results are similar to those of \cite{joshi22}.
Our PWN model here intends to test their conclusions after relaxing assumptions or adding additional physical details.

A recent paper by \cite{tibet21} proposed a hadronic origin of the LHAASO source based on spatial proximity of a molecular cloud (MC) with the gamma ray centroid of the source. However, as pointed out by \cite{breuhaus22}, the associated SNR is quite old to produce the observed hard gamma ray spectrum at the highest energies. Consequently, a hadronic scenario from a SNR+MC association would need a peculiar modeling to explain the observed UHE gamma ray spectrum observed from the source. A novel approach was explored in \cite{desarkar22} to explain the hadronic origin of LHAASO J1908+0621 and perhaps, a similar approach may play a role here as well.


\section{Brief description of the model}\label{sec2}

For this work, we have used the code \texttt{TIDE}, for which earlier applications can be found in, e.g., \cite{torres14,martin16,torres18}. 
%
%
%
The numerical code solves the evolution of leptonic pair distribution in the PWN, as a function of Lorentz factor $\gamma$ at time $t$ described by the equation,
\begin{equation}\label{eq1}
    \frac{\partial N(\gamma, t)}{\partial t} = Q(\gamma, t) - \frac{\partial}{\partial \gamma} [\dot{\gamma}(\gamma, t) N(\gamma, t)] - \frac{N(\gamma, t)}{\tau(\gamma, t)}
\end{equation}
The left hand side of equation \ref{eq1} describes the variation of lepton distribution in time. The first term on the right hand side is the lepton injection function Q($\gamma$, t), which is usually assumed as a broken power law 
\begin{equation}
\label{eq2}
\begin{split}
Q(\gamma, t) = Q_0(t)
\begin{cases}
(\gamma/\gamma_b)^{-\alpha_1}&  \gamma \le \gamma_b,\\
(\gamma/\gamma_b)^{-\alpha_2} &  \gamma > \gamma_b,\\
\end{cases}              
\end{split}
\end{equation} 
The second and third terms on the right hand side take into account radiative losses such as synchrotron, IC, Bremsstrahlung, adiabatic losses or heating, as well as escape of the particles (we assume Bohm diffusion) respectively (see \cite{martin12} for the incorporated formulae).
The normalization factor $Q_0$(t) is calculated using the spin down luminosity L(t) of the pulsar through the equation,
$
        (1 - \eta)L(t) = \int^{\gamma_{max}}_{\gamma{min}}\gamma m_e c^2 Q(\gamma, t) d\gamma
$
        where $\eta$ is the fraction of spin down power that goes on to power the magnetic field of the PWN. The magnetic field varies in time as a result of the balance between the magnetic field power and adiabatic gains or losses of the field due to the contraction or the expansion of the PWN \citep{torres13}. 
The reverberation phase of the PWN included in the model, has been considered to be the same as that in \cite{torres17}. The variation of the PWN radius is calculated by taking into account the age, SNR explosion energy, ambient medium density, expansion velocity of the nebula and pressure profiles of SNR at the position of the PWN. The model takes into account the change in pressure profiles depending on whether the PWN shell is surrounded by unshocked ejecta (i.e. R $<$ R$_{RS}$), or the shocked ejecta (i.e. R$_{RS}$ $<$ R $<$ R$_{SNR}$), where R$_{RS}$ and R$_{SNR}$ are the radii of the SNR reverse shock and the SNR respectively. After reverberation, an assumed Sedov expansion follows when the PWN reaches the pressure of the SNR. For more details, please see \cite{martin16,torres17,torres18}. For a more detailed accounting of the reverberation period, we need to account for the fact that the 
ejecta pressure is not constant \citep{Bandiera:2020, Bandiera:2022}. This is beyond the scope of this work, as for the moment, a prescription to deal with this fact
in the context of radiative PWN is unavailable. Given the likely young age of the source, a PWN would be 
before the time of largest compression (see below), 
and thus we expect the current approach to be acceptable.

\section{Results}\label{sec3}

\subsection{Braking index and true age exploration}\label{subsec_1}

In the works of \cite{joshi22} and \cite{yu22} the true age and the braking index were chosen to be 7000 years and 3, respectively.   Since the choice of the true age is essential for a time evolutionary model of PWNe, we explore other values here as well. 
On the other hand, if the pulsar is assumed to be a dipole in vacuum, and considered to be spinning down by emitting only magnetic dipole radiation (MDR), then the corresponding braking index associated with the spin down is calculated to be 3 \citep{manchester77}. 
Alternatively, the spin down of a pulsar driven entirely by a particle wind would result in a braking index of 1 \citep{michel69, manchester85}. A combination of magnetic dipole radiation and wind braking would result in a braking index with a value in between 1 and 3 \citep{archibald16}. Most of the observed pulsar braking indices falls within this range (\cite{espinoza11, pons12}, and references therein). Although there are exceptions, for example PSR J1640--4631, which has a braking index of n = 3.15 $\pm$ 0.03 \citep{archibald16} and PSR J1734--3333 has a braking index of n = 0.9 $\pm$ 0.2 \citep{espinoza11b}. 
For this work, we only consider the braking index range $1 < n < 3$. Since the choice of the braking index affects the characteristics of the pulsar spin down, it is also important to explore whether the variation of $n$ affects the SED.

\begin{figure*}[htp]

\centering
\includegraphics[width=.33\textwidth]{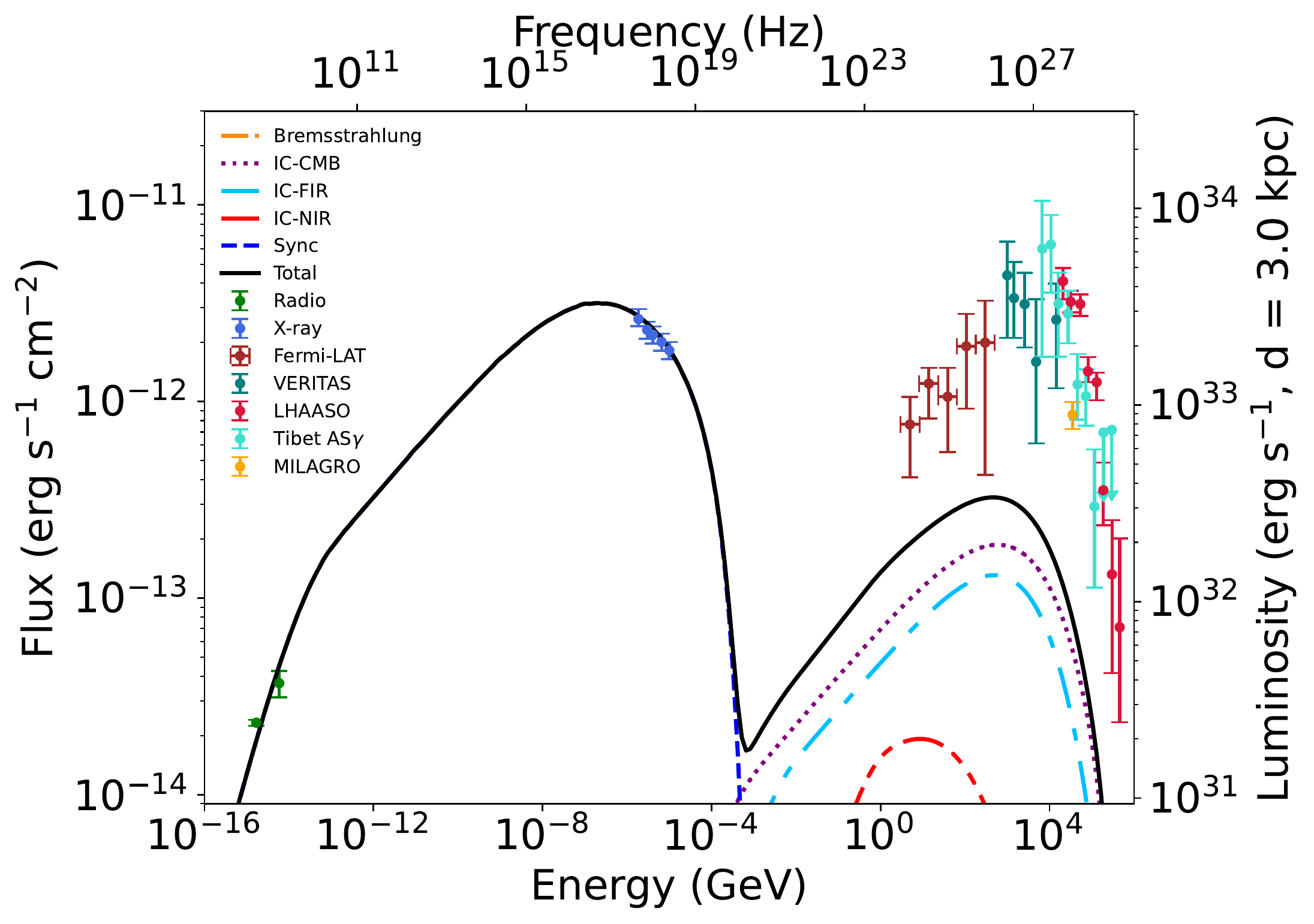}
\includegraphics[width=.33\textwidth]{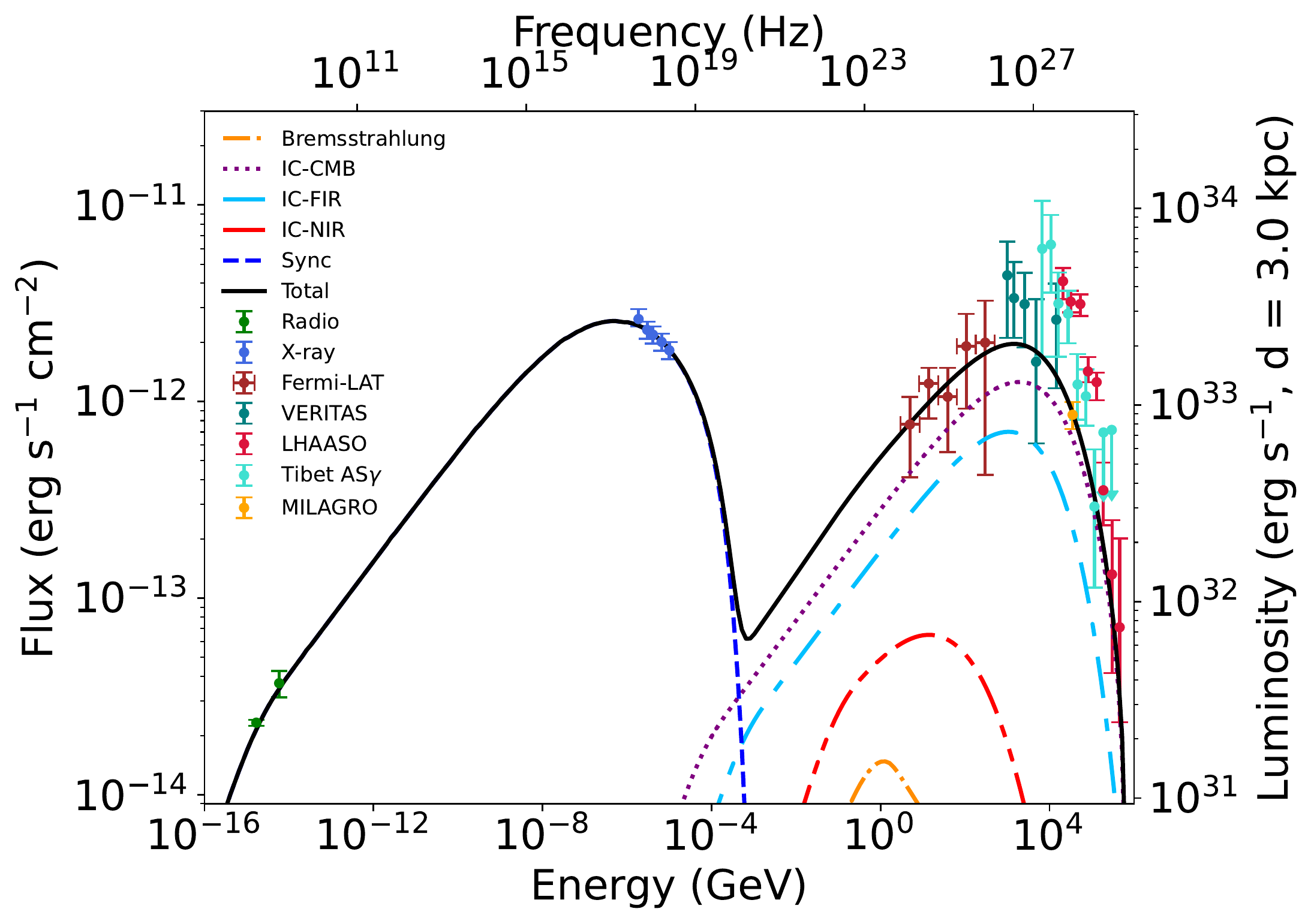}
\includegraphics[width=.33\textwidth]{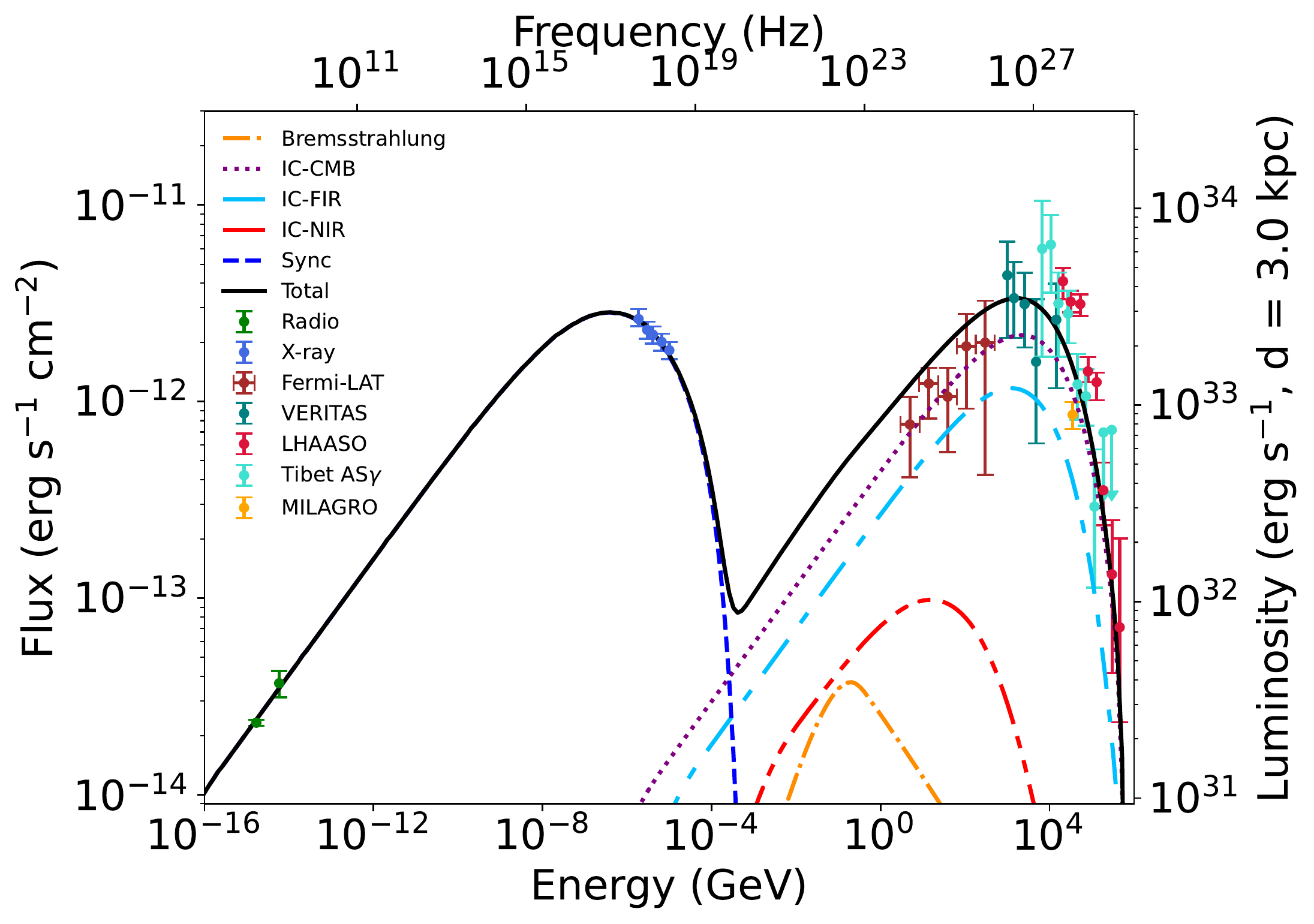}
\includegraphics[width=.33\textwidth]{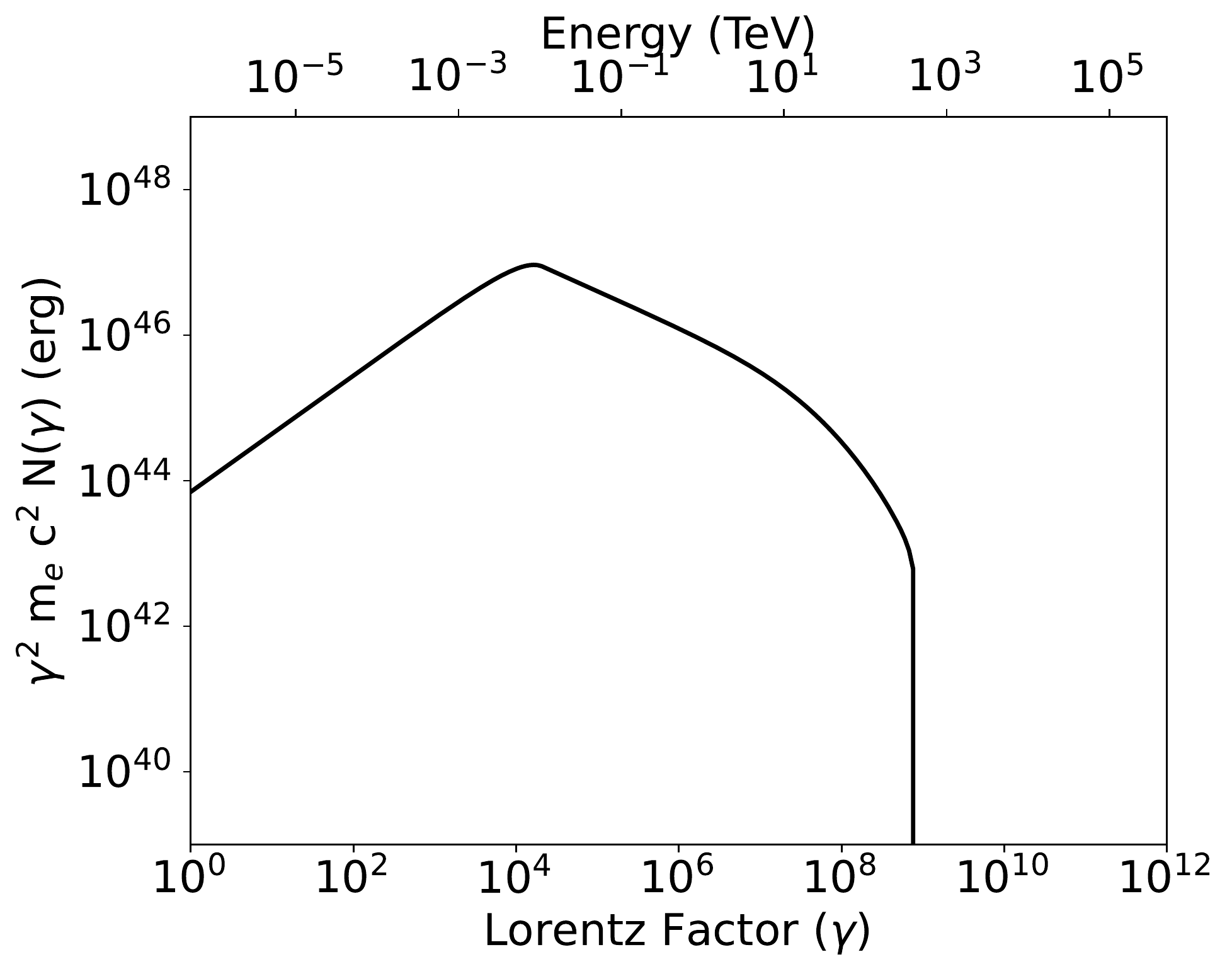}
\includegraphics[width=.33\textwidth]{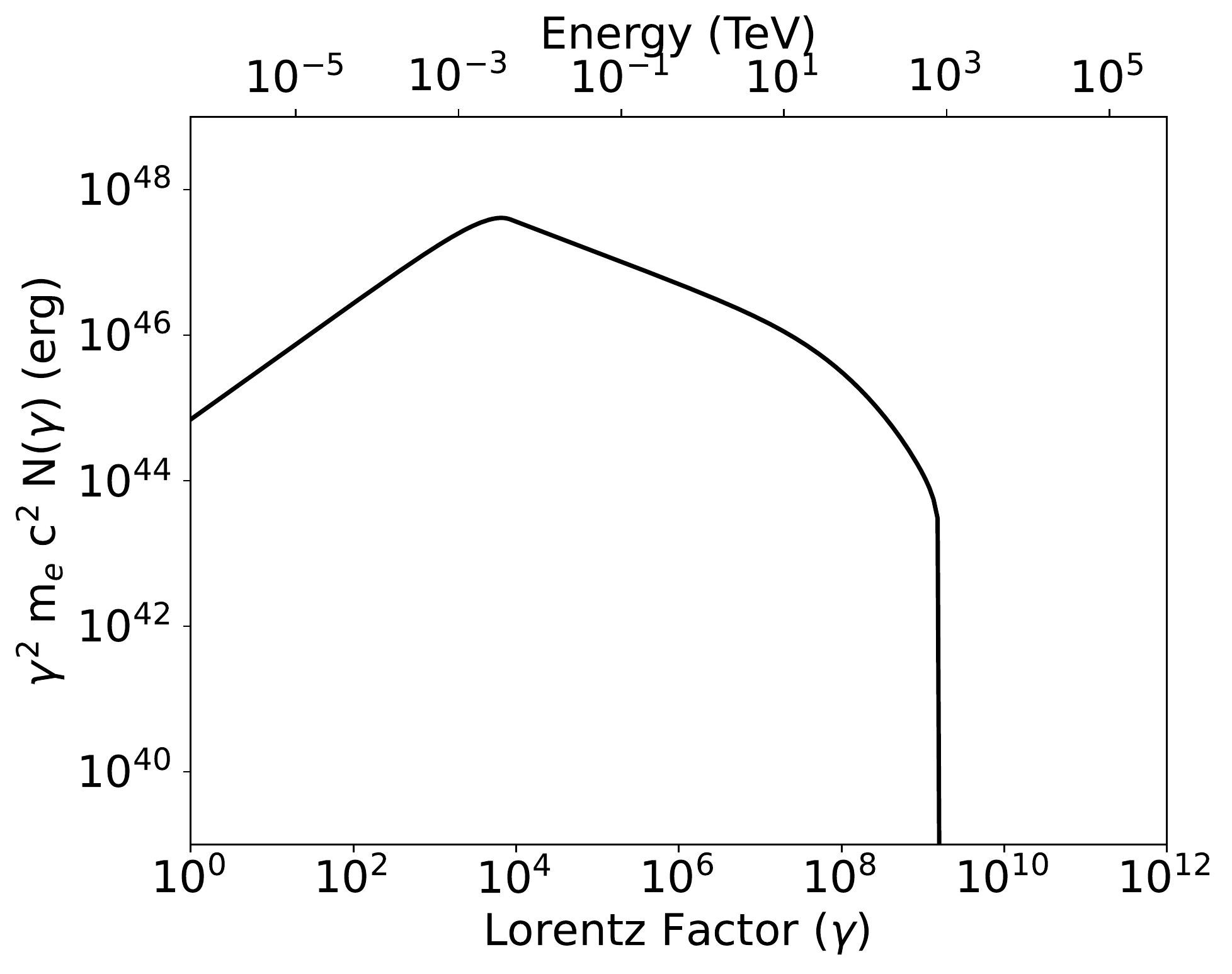}
\includegraphics[width=.33\textwidth]{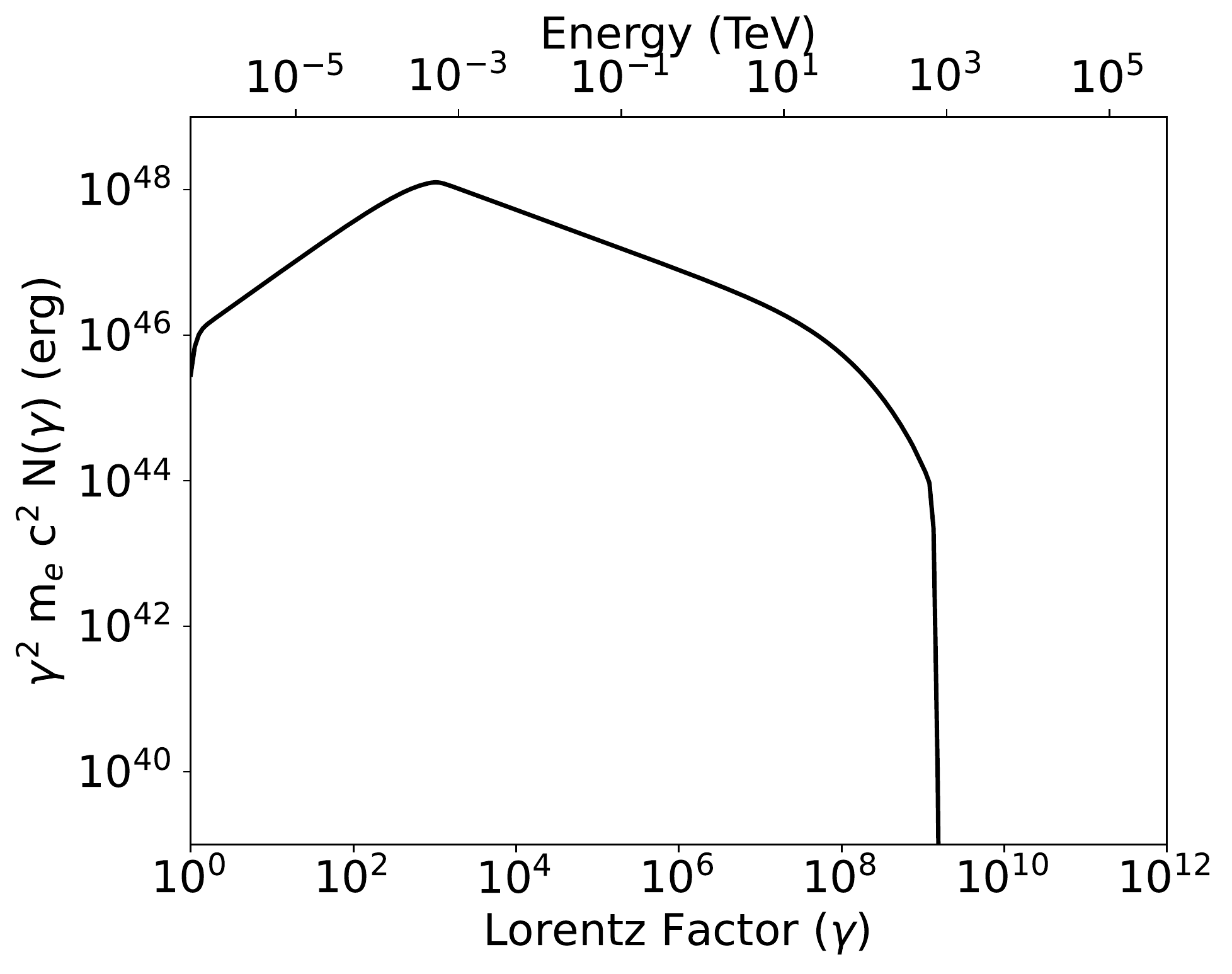}
\includegraphics[width=.33\textwidth]{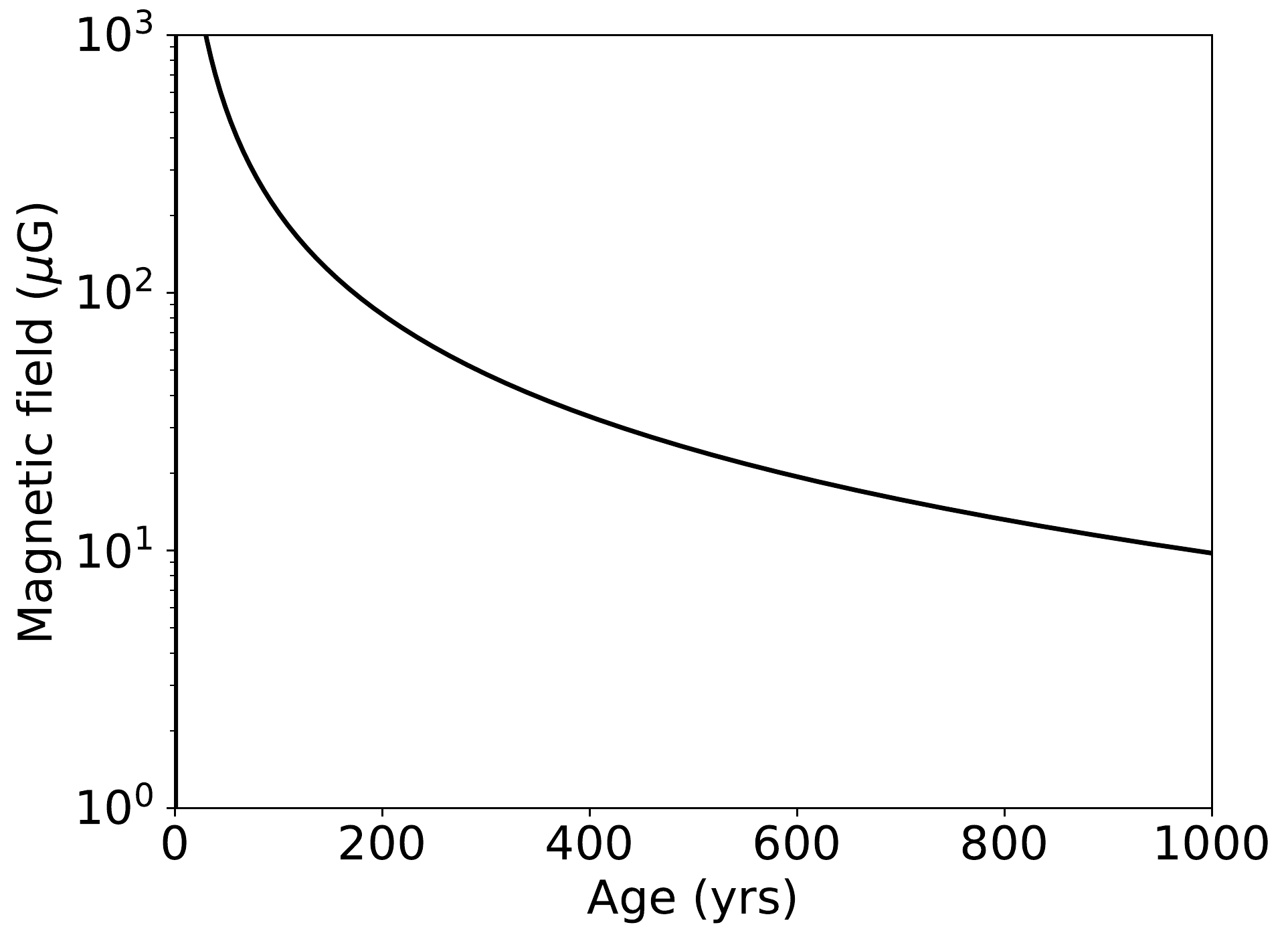}
\includegraphics[width=.33\textwidth]{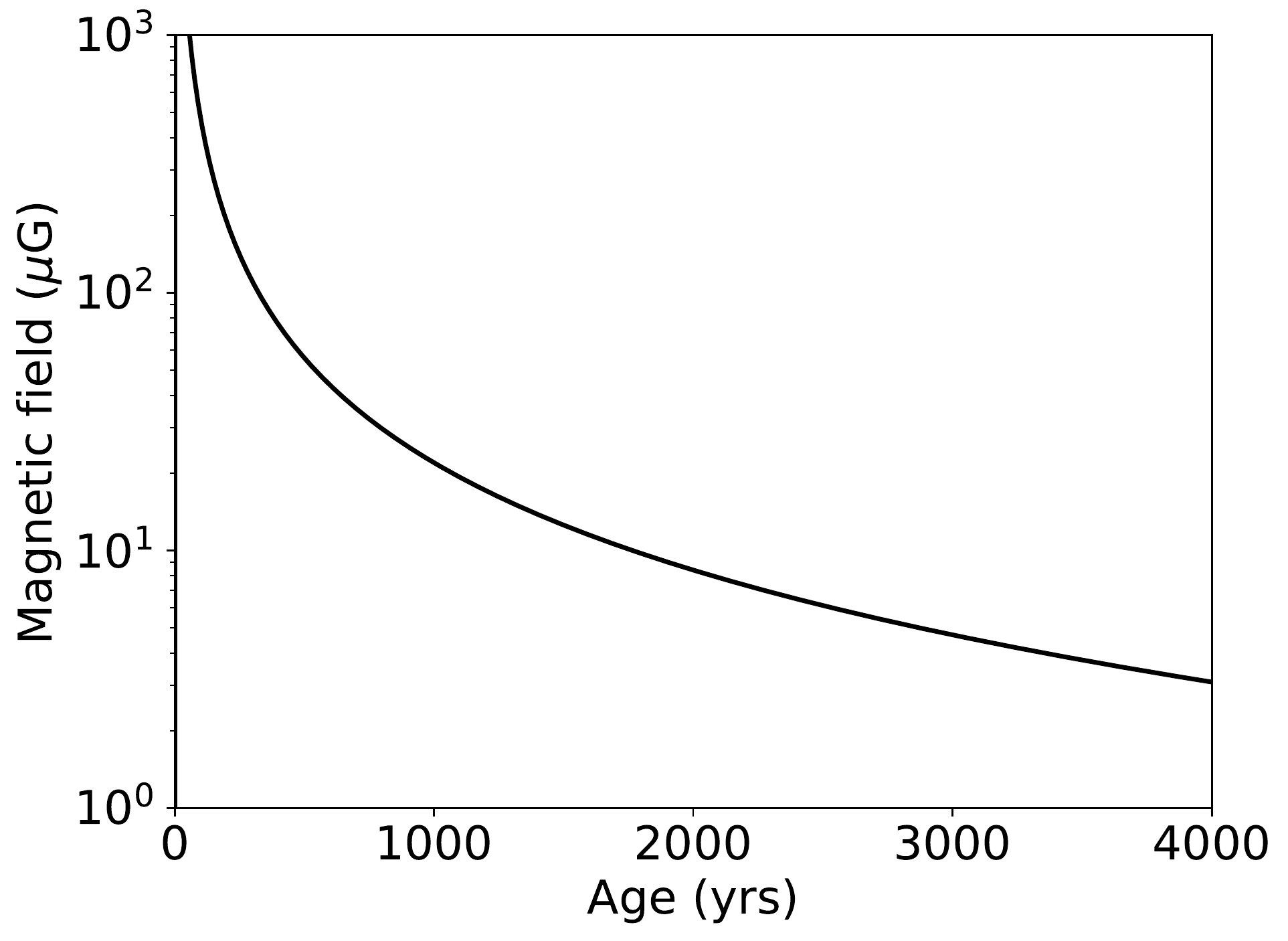}
\includegraphics[width=.33\textwidth]{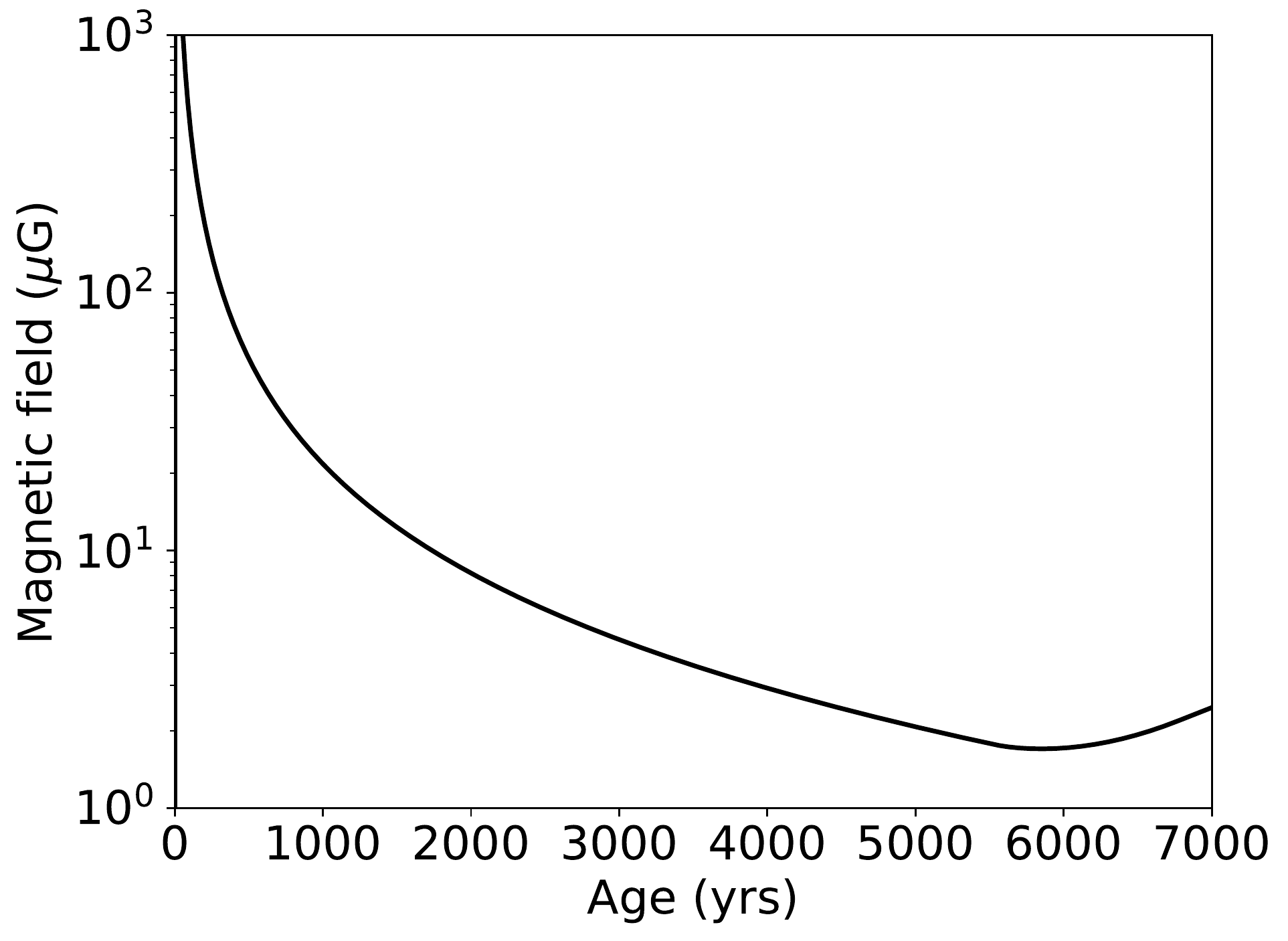}
\caption{The LHAASO J2226+6057 MWL SEDs along with the calculated model flux (top row), present lepton spectrum (middle row) and magnetic field evolution with time (bottom row) are given for $t_{age}$ = 1000 years (left column), 4000 years (middle column) and 7000 years (right column), for a fixed braking index n = 2.5. The radio data (green) and X-ray data (royalblue) are taken from \cite{pineault2000} and \cite{fujita21} respectively. Fermi-LAT (brown), VERITAS (teal), Tibet AS$\gamma$ (turquoise), MILAGRO (orange) and LHAASO (crimson) data are taken from \cite{xin19}, \cite{acciari09}, \cite{tibet21}, \cite{milagro09} and \cite{cao21}, respectively.}
\label{fig1}
\end{figure*}

\begin{figure*}[htp]
\centering
\includegraphics[width=.5\textwidth]{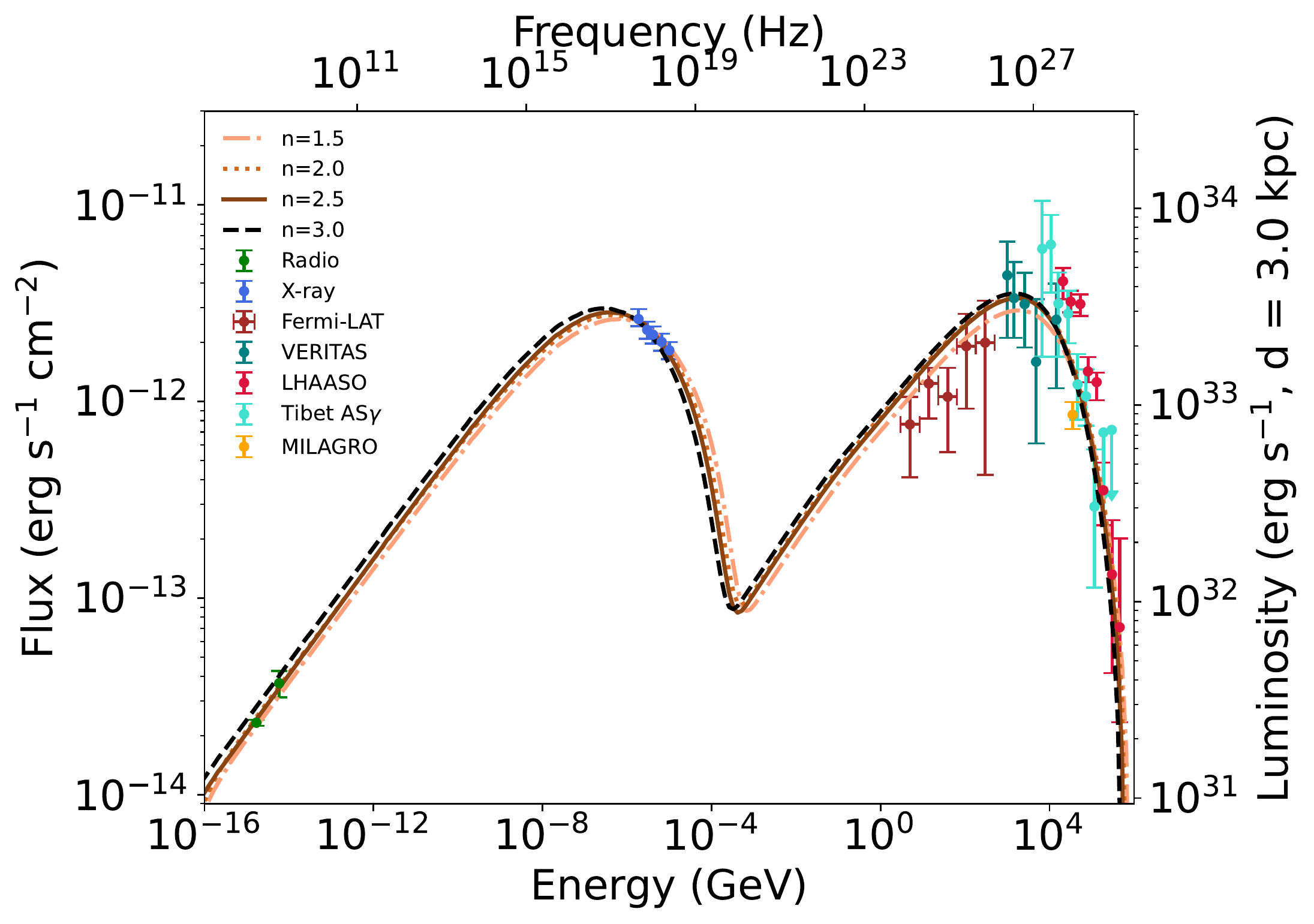}
\caption{MWL SEDs of LHAASO J2226+6057 for $n = 1.5, 2.0, 2.5$, and 3.0. The color scheme of the data points is same as given in Figure \ref{fig1}.}
\label{fig2}
\end{figure*}

To explore the effect of $n$ and $t_{age}$ on the MWL SED, we have chosen $t_{age}$ = 1000, 4000 and 7000 years, and $n = 1.5, 2.0, 2.5$ and 3.0. Given the characteristic age $\tau_c$ $\approx$ 10.5 kyr and present day spin down luminosity $L(t_{age}) \approx 2.25 \times 10^{37}$ erg/s, the initial spin down age ($\tau_0$) and initial spin down luminosity ($L_0$) were calculated using the relations,
\begin{equation}
\label{eq4}
    L (t) = L_0 \left(1 + \frac{t}{\tau_0}\right)^{-\frac{n+1}{n-1}},
\end{equation}
and,
\begin{equation}
\label{eq5}
    \tau_0 = \frac{2 \tau_c}{n - 1} - t,
\end{equation}
for a specific choice of $n$ and $t_{age}$. 
%
We also assume the values of the following parameters are the same throughout the work; minimum Lorentz factor $\gamma_{min}$ = 1, SN explosion energy E$_{SN}$ = 10$^{51}$ erg, interstellar medium (ISM) ambient density $\rho_{ISM}$ = 0.1 cm$^{-3}$, SNR core density index = 0, SNR envelope density index = 9 (we assume a type II SN as the progenitor, as in \cite{chevalier92, blondin01, gelfand09}), PWN adiabatic index = 1.333 and SNR adiabatic index = 1.667, containment factor $\epsilon$ (ratio of the Larmor radius of particles to the radius of the termination shock) = 0.5 and magnetic compression ratio $\kappa$ = 3 (strong shock condition). 
The ejected mass of the progenitor SNR was also fixed at $M_{ej}$ = 8$M_{\odot}$ for the $n-t_{age}$ exploration. 
For the target radiation fields for the IC interaction, we consider the cosmic microwave background, 
Far-Infrared (FIR) and Near-Infrared (NIR) radiation fields at the source position with temperatures of $T_{FIR}$ = 25 K and $T_{NIR}$ = 5000 K, respectively. The energy densities associated with FIR and NIR field were taken from \cite{porter06} and fixed at those values for $n-t_{age}$ exploration. It has been observed (see \cite{torres14}) that almost always the energy densities of FIR/NIR fields required to explain the IC emission from PWNe, differ from the interstellar radiation fields reported by \cite{porter06}. 
\cite{breuhaus21} has also argued that an enhancement of the radiation field is needed to explain the hard gamma ray spectrum observed at highest energies. 
However, since the aim of this subsection is to compare the effect of $n$ and $t_{age}$ on the computed SED, we have fixed the energy densities of FIR/NIR fields at the values given by \cite{porter06}, i.e. $\omega_{FIR}$ = 0.29 eV cm$^{-3}$ and $\omega_{NIR}$ = 0.45 eV cm$^{-3}$, so there are less free parameters at play.
The maximum Lorentz factor of the lepton distribution was fixed by the most restrictive condition between the synchrotron limit \citep{dejager96} or gyroradius limit \citep{dejager09} during the evolution. For a specific choice of $n$ and $t_{age}$, we vary the injection function parameters, i.e. low energy index $\alpha_1$, high energy index $\alpha_2$, energy break $\gamma_b$, as well as the magnetic fraction $\eta$, to describe the MWL SED of the source.

First, we fix the braking index at the value of $n = 2.5$, and consider three different cases of $t_{age}$. The MWL SED along with the computed model flux, injected lepton spectrum and magnetic field at the considered $t_{age}$ are shown in Figure \ref{fig1}. From the figure, it can be seen that the fit corresponding to $t_{age}$ = 7000 years is the best of the three ages considered.
For $t_{age}$ = 1000 years and $t_{age}$ = 4000 years, the X-ray and radio data can be adequately explained. However, for $t_{age}$ = 1000 years, the IC emission is not significant enough to explain the high-energy data. Also for $t_{age}$ = 4000 years, the IC emission only partially explains it. From this exploration, it is apparent that the age of the PWN lies within 4000 and 7000 years and we consider $t_{age}$ = 7000 years to compare with \cite{joshi22} and \cite{yu22}, albeit these authors did not explore other options. We also discuss the results of considering the true age as a free parameter at a later subsection.

After limiting the age of the system, we move on to explore the effect of braking index on the evolution of the source. Assuming a fixed $t_{age}$ = 7000 years, we change the braking index, and try to describe the MWL SED. $L_0$ and $\tau_0$ have been calculated using equations \ref{eq4} and \ref{eq5}. We again change $\alpha_1$, $\alpha_2$, $\gamma_b$ and $\eta$, similarly to the previous case. 
The calculated spectra for $n = 1.5, 2.0, 2.5,$ and 3.0 are given in Figure \ref{fig2}.  $n$ does not significantly affect the computed SEDs and we have selected $n = 2.5$ for further study. 

\subsection{$\chi^2$ fitting of the MWL SED}\label{subsec_2}

\begin{figure*}[htp]
\centering
\includegraphics[width=.60\textwidth]{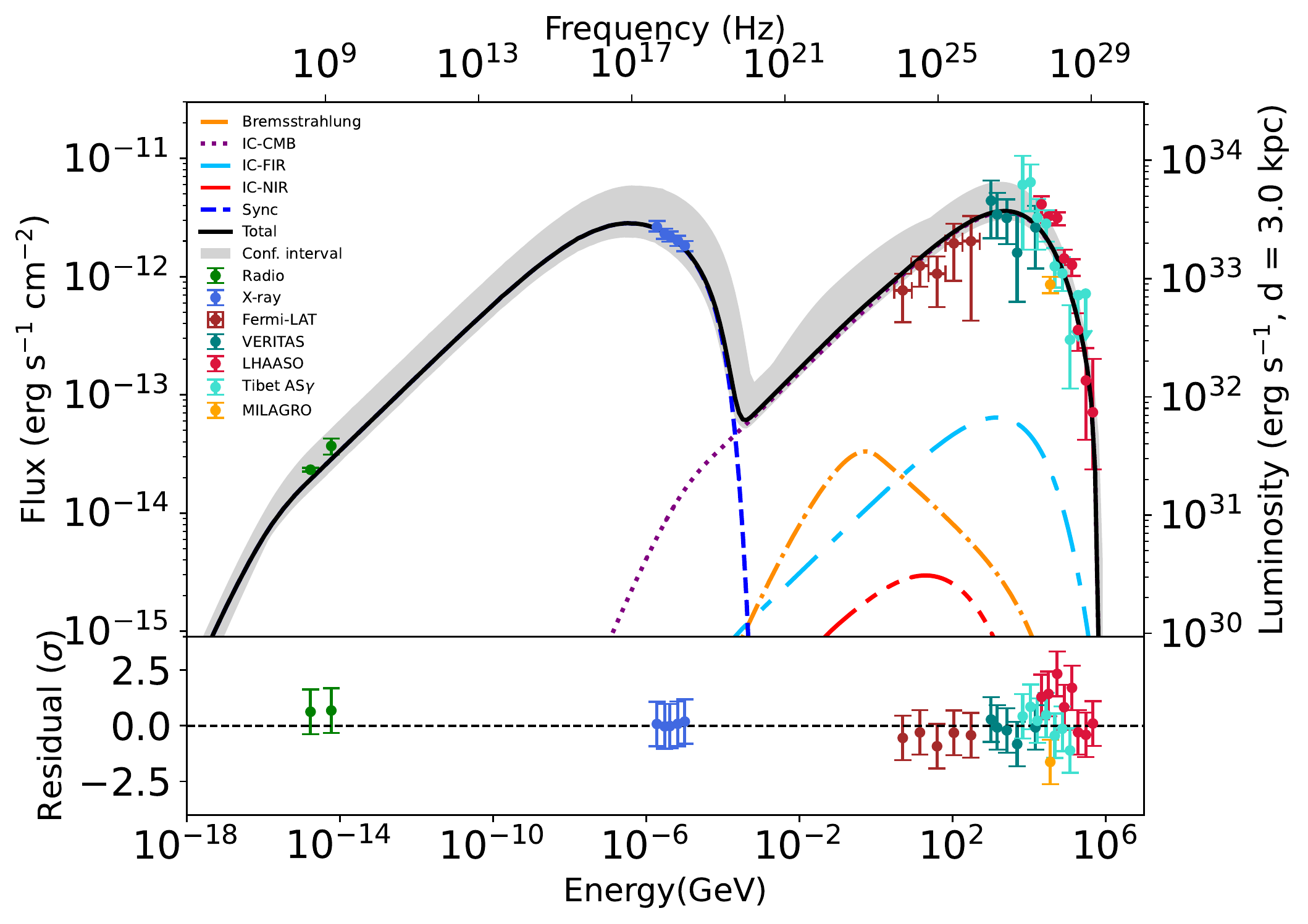}
\includegraphics[width=.47\textwidth]{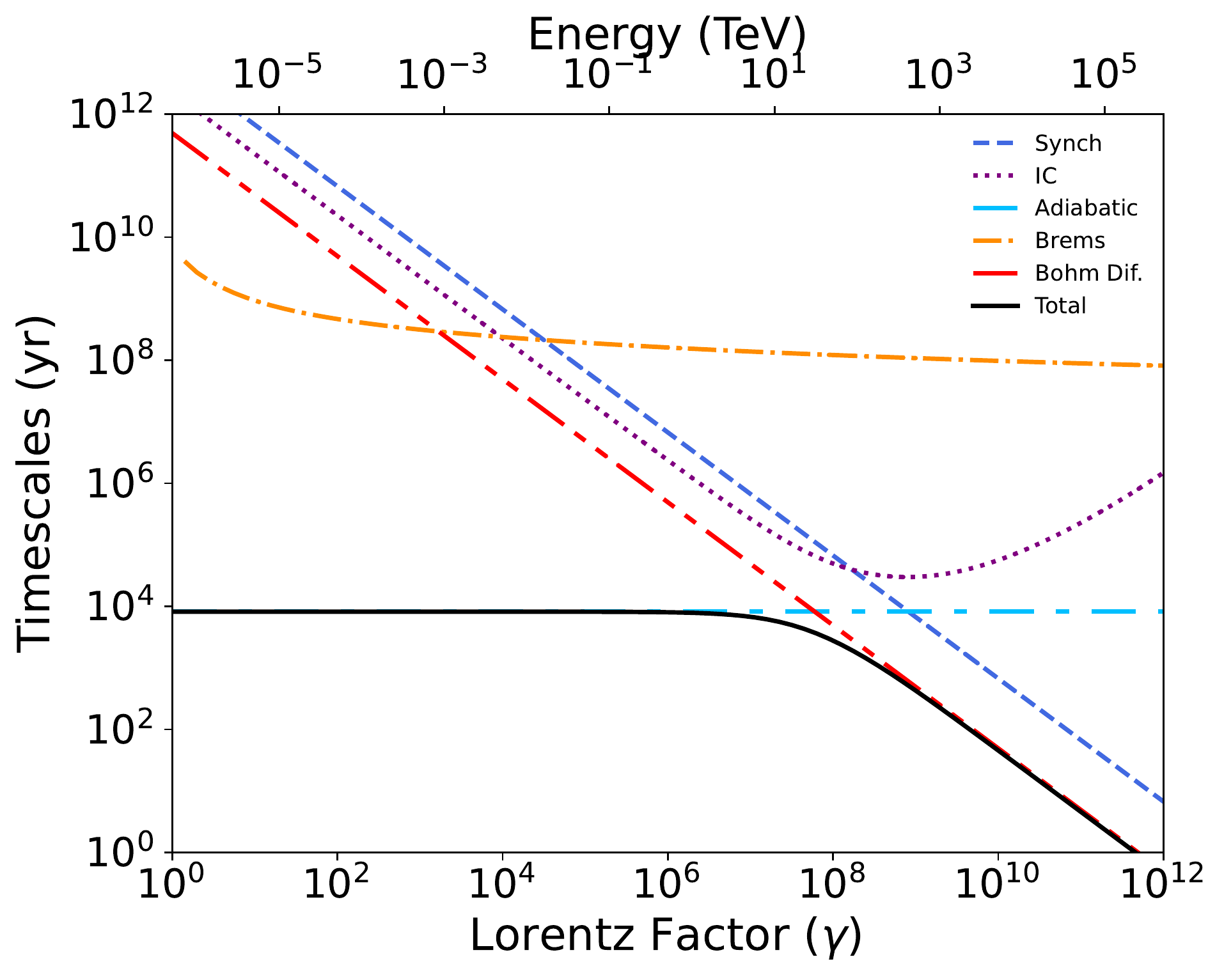}
\includegraphics[width=.47\textwidth]{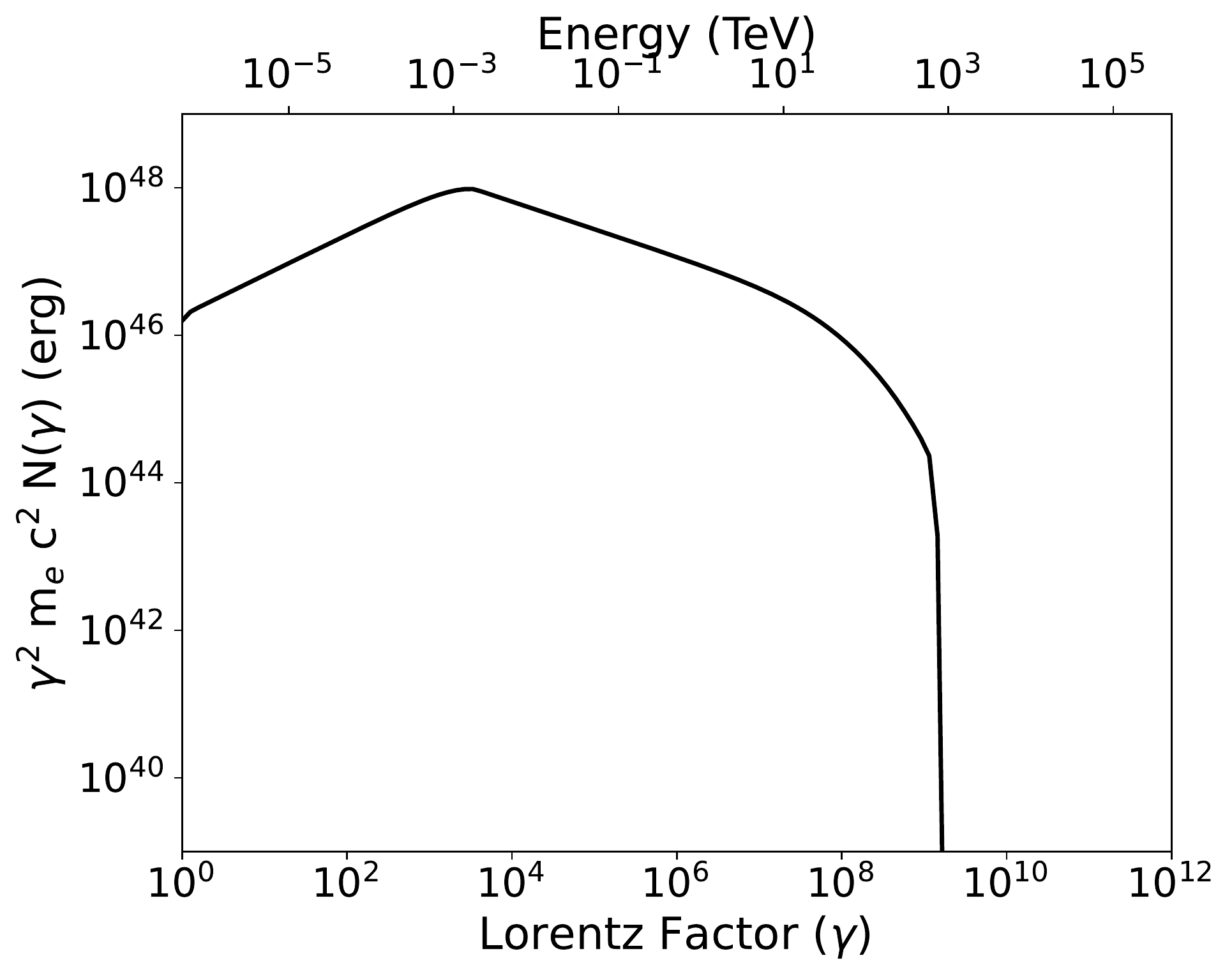}
\includegraphics[width=.47\textwidth]{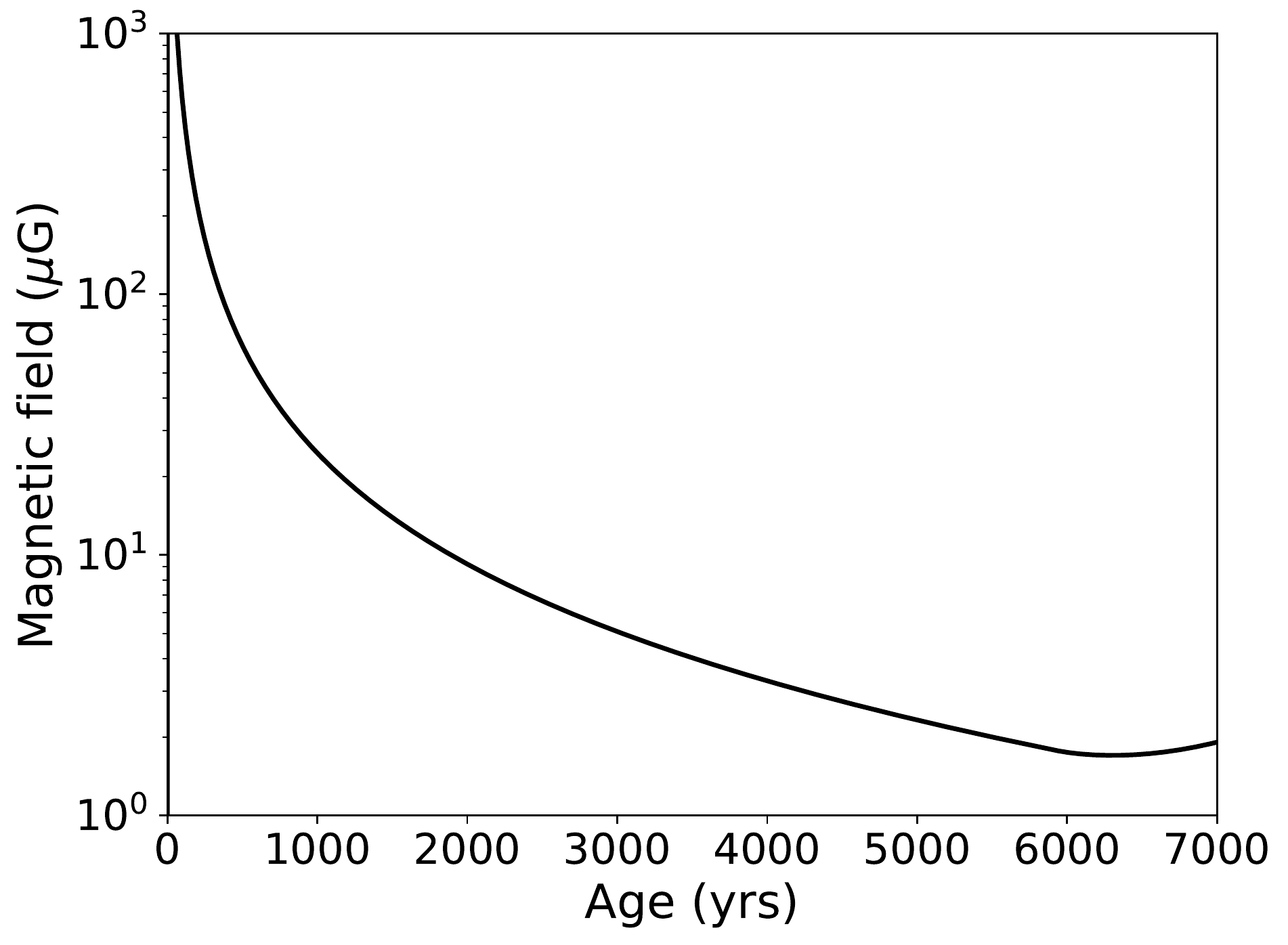}
\includegraphics[width=.47\textwidth]{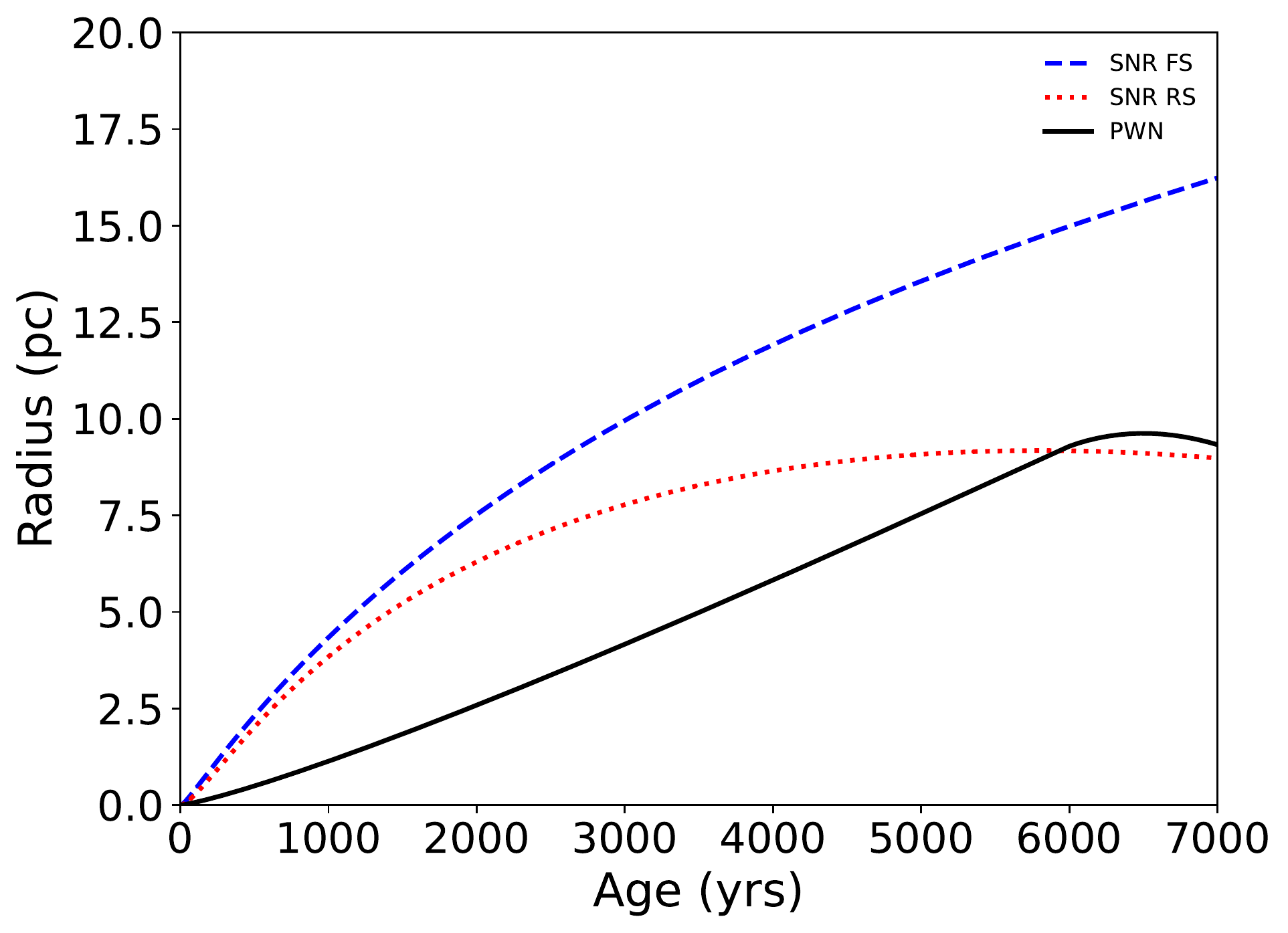}
\caption{The calculated best-fit MWL spectrum is shown at the top row, along with the MWL data points. The color scheme of the data points are same as that in Figure \ref{fig1}. In the bottom pannel of the figure, the residuals are also plotted. The color scheme of the residuals are same as the data points. The middle row shows the timescales of radiative losses, adiabatic losses and the escape of particles considered in the model (left) and the injected lepton spectrum at the present age (right). Also the time evolution of the magnetic field (left), and SNR forward shock, SNR reverse shock and PWN radius (right) are given in the bottom row.}
\label{fig3}
\end{figure*}

\begin{figure*}[htp]
\centering
\includegraphics[width=.49\textwidth]{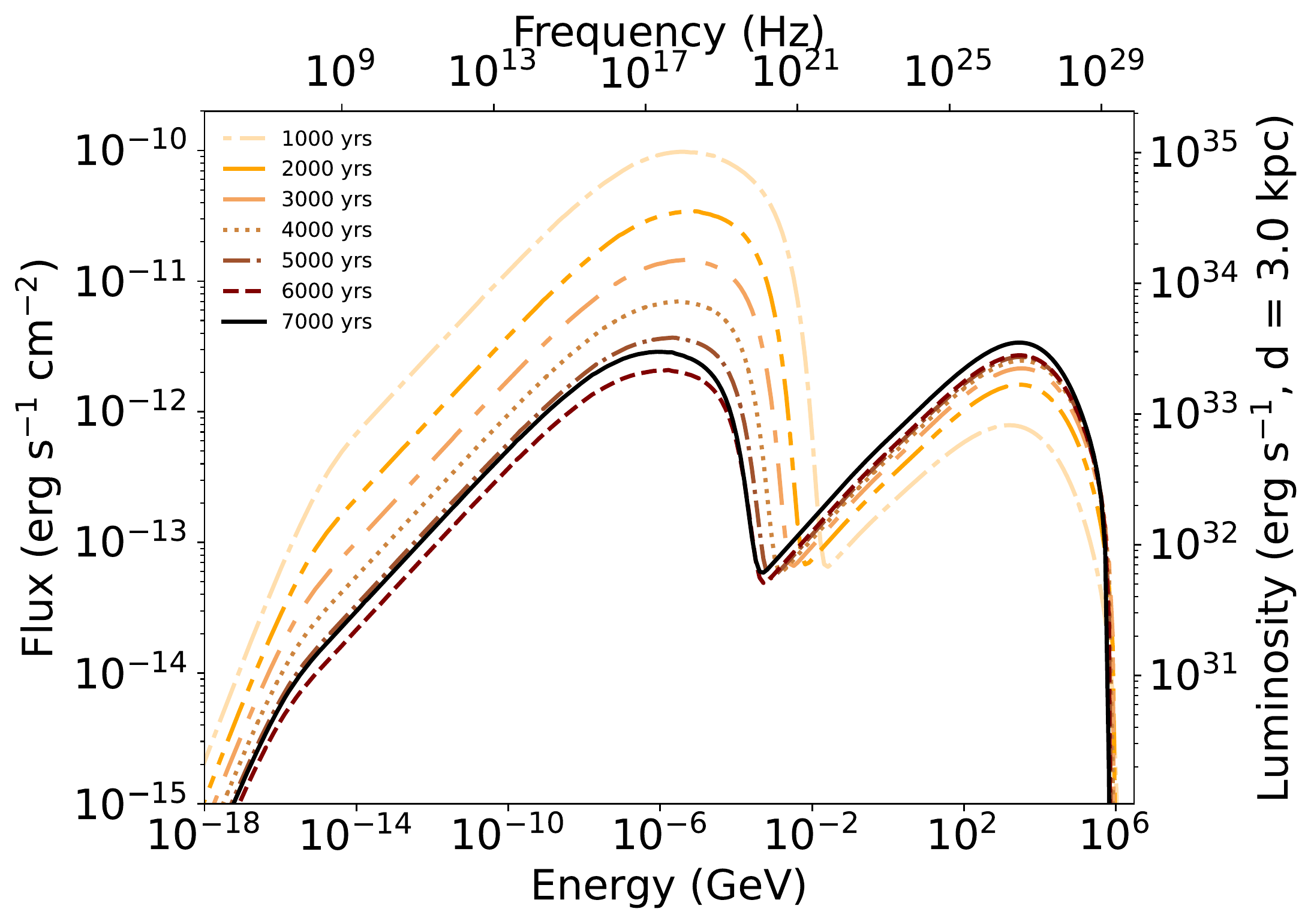}
\includegraphics[width=.49\textwidth]{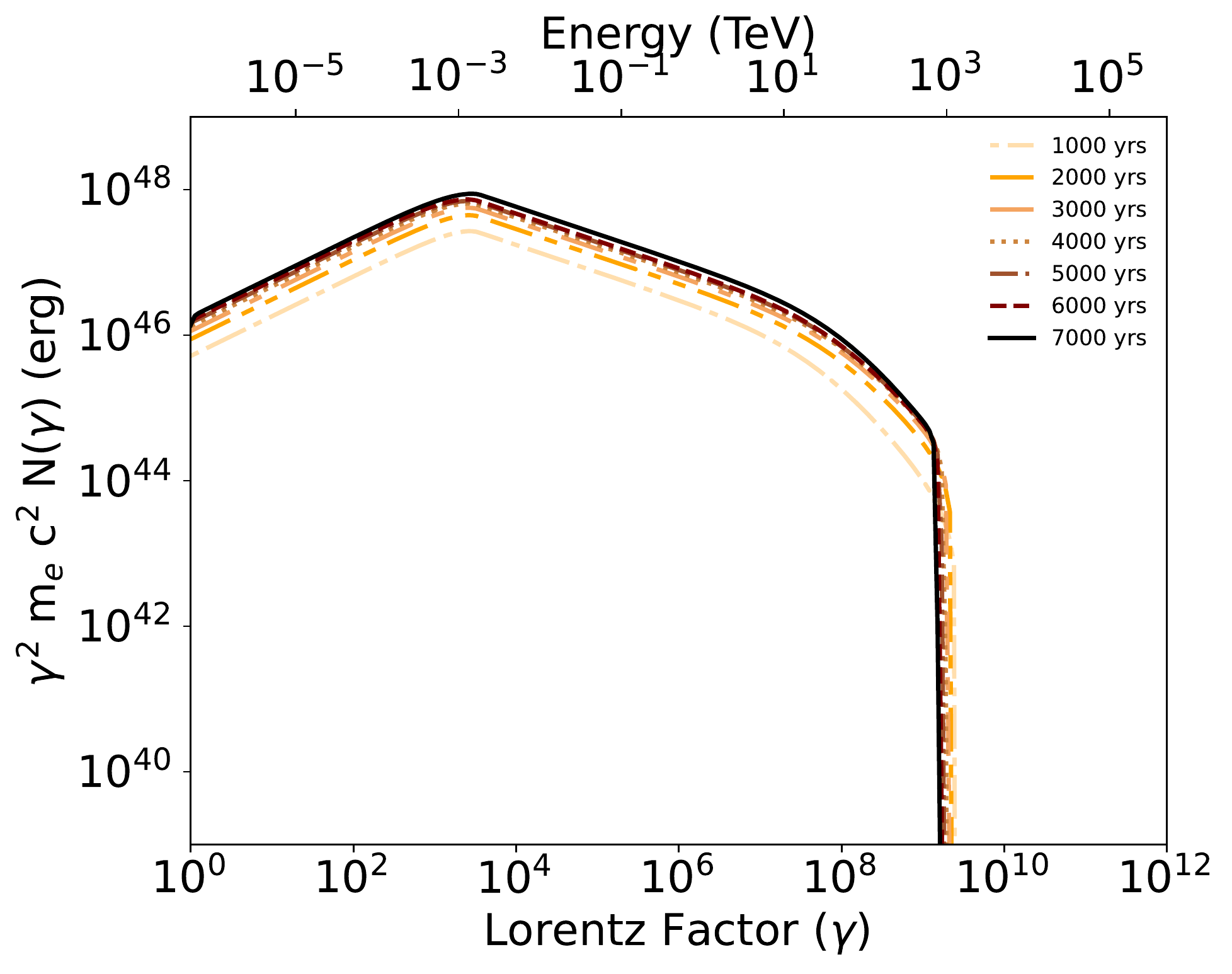}
\caption{The time evolution of calculated MWL spectrum (left), as well as the injected lepton spectrum (right) are shown in the figure, assuming the parameters given in Table \ref{tab1}.}
\label{fig4}
\end{figure*}

We have considered typical PWN parameters as initial input, and, using the \texttt{TIDEFIT} code \cite{martin22}, subsequently varied them to solve equation \ref{eq1} and compute the best $\chi^2$ fitted model spectrum. 
To find the best-fit spectra, we have used $t_{age}$ = 7000 years and $n = 2.5$. Similar to the previous discussion, $L_0$ and $\tau_0$ have been fixed at the value calculated by equation \ref{eq4} and \ref{eq5}. 
The PWN parameters that have been fixed already in the above discussion are also fixed in this calculation as well, except for the cases of ejected mass and FIR/NIR energy densities (between 0.01 eV cm$^{-3}$ to 5 eV cm$^{-3}$). Since the ejected mass of the progenitor SNR directly affects the size, as well as the magnetic field of the PWN, it was left free within the typical ejected mass range of 7M$_{\odot}$ to 15M$_{\odot}$.
Apart from these changes, similarly to the previous discussion, $\alpha_1$, $\alpha_2$, $\gamma_b$ and $\eta$ were left free to vary to find the best-fit MWL spectrum. The parameters used in the model are reported in Table \ref{tab1}, in which we have divided the parameters among measured or assumed, derived, and fitted values. The resulting plots are given in Figure \ref{fig3}. The time evolution of the calculated MWL spectrum and injection spectrum are given in Figure \ref{fig4}.  

From the top row of Figure \ref{fig3}, it can be seen that the computed MWL spectrum using the model along with the 1$\sigma$ confidence interval, matches well with the observed MWL data. The goodness of fit can also be seen from the bottom residual plot. The systematic uncertainty associated with the model is  0.32, and the $\chi^2$/D.O.F. for the given fit is 35.65/30. Also from the figure, as well as from Table \ref{tab1}, it can be seen that FIR and NIR radiation fields do not contribute to the IC emission needed to explain the VHE-UHE data, rather it was found the CMB is most likely solely responsible as the target photon field required. 
A recent work by \cite{wilhelmi22} also considered CMB photons as the most relevant target for IC scattering. 
%
However, such result is somewhat different from what was assumed by \cite{joshi22}, in which they have considered radiation fields 1.5-3.0 times that of CMB to fit the data. 

From the bottom row of Figure \ref{fig3} and Table \ref{tab1}, it can be seen that the PWN would be  very extended at the present age according to the model, and concurrently the associated magnetic field would be very low, close to the Galactic average value. 
The large radius of the PWN may agree with the large extension measured by LHAASO, but it contradicts with the radio and X-ray sizes observed for the Boomerang PWN \citep{halpern01, Halpern_2001b}, as these sizes are much smaller compared to the calculated PWN radius. The required magnetic field to fit the data is also uncomfortably low (we discuss this below), and begs the question how are the particles confined in such diluted PWN. 

\begin{table*}
\caption{Physical parameters used by and resulting from the fit. The bracketed terms in the fitted parameters section signify the lower and  upper bounds of 1$\sigma$ confidence interval respectively.}             
\label{tab1}      
\centering          
\begin{tabular}{l l l}      
\hline\hline       
                      
Definition & Parameter & Value\\ 
\hline
Measured or assumed parameters: & & \\
\hline
   Age & $t_{age}$ [kyr] & 7\\  
   Characteristic age & $\tau_c$ [kyr] & 10.5 \\
   Braking index & n & 2.5\\
   Present day spin down luminosity & L($t_{age}$) [erg s$^{-1}$] & 2.25 $\times$ 10$^{37}$ \\
   Distance & D [kpc] & 3 \\
   Minimum energy at injection & $\gamma_{min}$ & 1 \\
   SN explosion energy & E$_{SN}$ [erg] & 10$^{51}$\\
   ISM density & $\rho_{ISM}$ [cm$^{-3}$] & 0.1 \\
   SNR core density index & w$_{core}$ & 0\\
   SNR envelope density index & w$_{env}$ & 9\\
   PWN adiabatic index & $\gamma_{PWN}$ & 1.333\\
   SNR adiabatic index & $\gamma_{SNR}$ & 1.667\\
   Containment factor & $\epsilon$ & 0.5\\
   Magnetic compression ratio & $\kappa$ & 3\\
   CMB temperature & T$_{CMB}$ [K] & 2.73\\
   CMB energy density & $\omega_{CMB}$ [eV cm$^{-3}$] & 0.25\\
   FIR temperature & T$_{FIR}$ [K] & 25\\
   NIR temperature & T$_{NIR}$ [K] & 5000\\
   \hline
   Derived parameters: & & \\
   \hline
   Initial spin down luminosity & L$_0$ [erg s$^{-1}$] & 1.13 $\times$ 10$^{38}$\\
   Initial spin down age & $\tau_0$ [kyr] & 7\\
   \hline
   Fitted parameters: & & \\
   \hline
   Energy break at injection & $\gamma_b$ & 3338.00 (2082.91, 10597.30)\\
   Low energy index at injection & $\alpha_1$ & 1.4522 (1.0000, 1.6432)\\
   High energy index at injection & $\alpha_2$ & 2.3727 (2.3316, 2.3890)\\
   Ejected mass & $M_{ej}$ [$M_{\odot}$] & 8.8927 (8.1735, 9.3202)\\
   Magnetic fraction & $\eta$ & 0.0033 (0.0026, 0.0060)\\
   FIR energy density & $\omega_{FIR}$ [eV cm$^{-3}$] & 0.0100 (0.0100, 0.4611)\\
   NIR energy density & $\omega_{NIR}$ [eV cm$^{-3}$] & 0.0100 (0.0100, 5.0000)\\
   \hline
   Resulting features: & &\\
   \hline
   PWN radius & R$_{PWN}$ (t$_{age}$) [pc] & 9.33\\
   SNR forward shock radius & R$_{FS}$ (t$_{age}$) [pc] & 16.23\\
   SNR reverse shock radius & R$_{RS}$ (t$_{age}$) [pc] & 8.98\\
   PWN magnetic field & B$_{PWN}$ (t$_{age}$) [$\mu$G] & 1.91\\
\hline
\hline
\end{tabular}
\end{table*}

\subsection{Possible impact of reverberation}\label{subsec_3}

From Figure $\ref{fig3}$, and from the resulting values given in Table \ref{tab1}, it can be seen that the radius of the SNR reverse shock is smaller than the PWN radius, which means that the reverse shock has just reached the position of PWN shell, at the onset of the reverberation phase of the PWN. If the age of the system is increased further from the considered age, the effect of reverberation will be apparent on the MWL spectrum. Since the PWN has just started to contract at the considered age of 7000 years, it is unlikely that it is heavily distorted.

\begin{figure*}[htp]
\centering
\includegraphics[width=.49\textwidth]{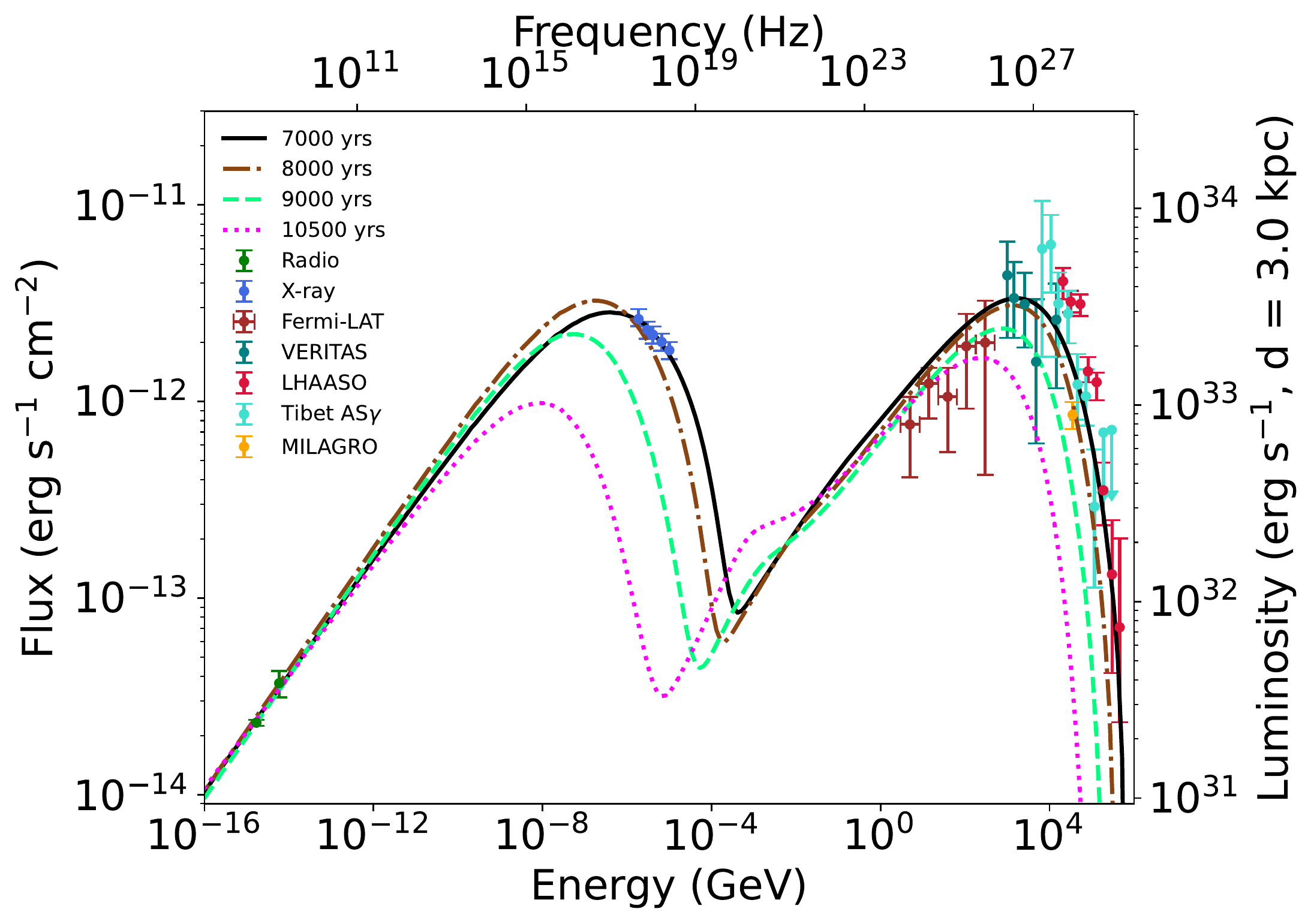}
\includegraphics[width=.49\textwidth]{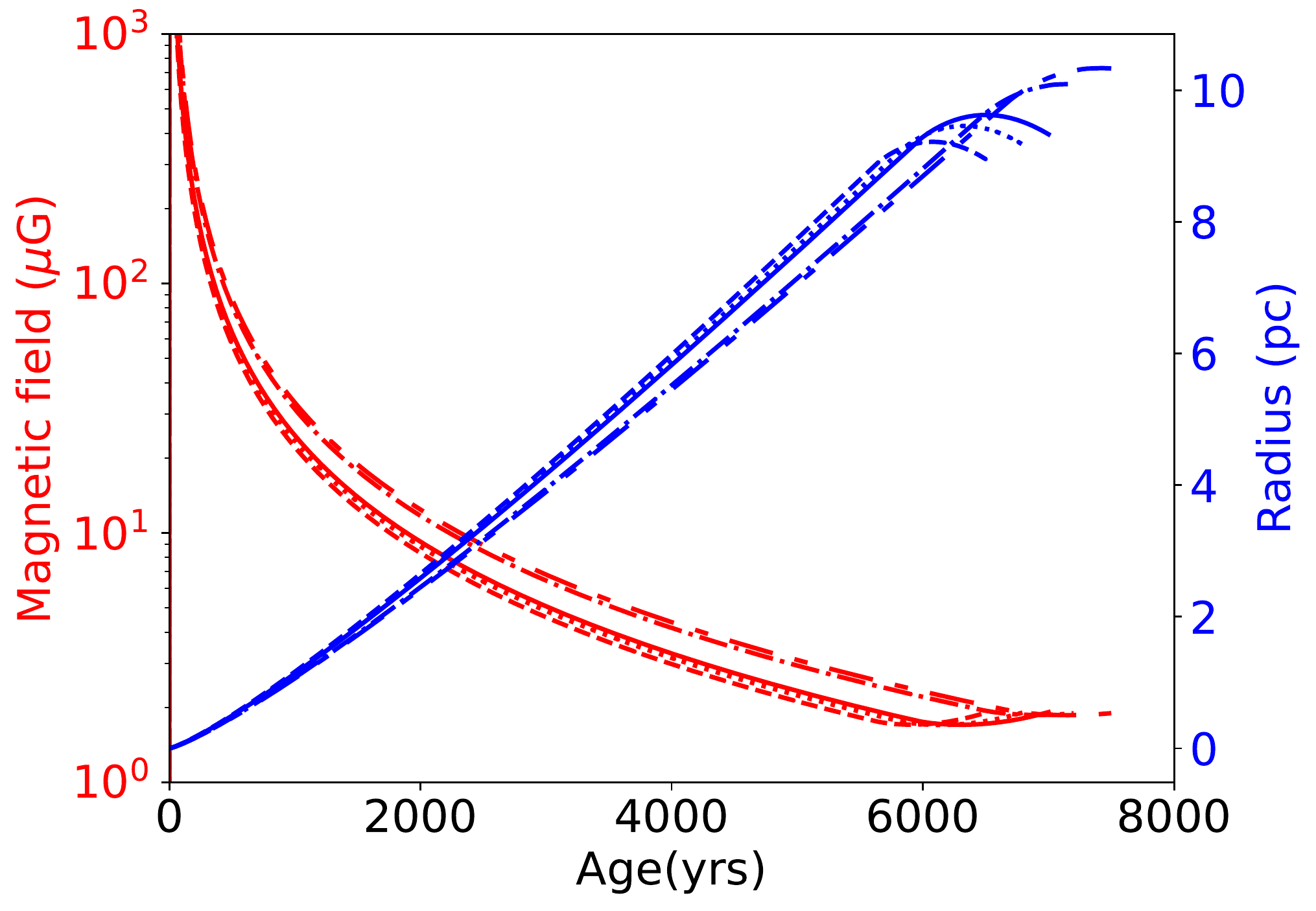}
\caption{Impact of reverberation on the PWN MWL spectrum with increasing age is shown (left). Time evolution of PWN magnetic field (red) and radius (blue) is shown (right) for $t_{age}$ = 6500 years (dashed), 6800 years (dotted), 7000 years (solid), 7200 years (dot-dashed) and 7500 years (long-dashed).}
\label{fig6}
\end{figure*}

We consider $t_{age}$ = 7000, 8000, 9000 and 10500 years to study the effect of reverberation on the MWL spectrum. The braking index was chosen to be 2.5, and $L_0$ and $\tau_0$ were calculated using equation \ref{eq4} and \ref{eq5} for a specific choice of $t_{age}$. 
Since we are only interested in comparing the effect of reverberation on the MWL spectrum based on the assumption of different ages of the PWN, we assume the values of \cite{porter06} for the FIR/NIR energy densities. Similarly to subsection \ref{subsec_1}, $\alpha_1$, $\alpha_2$, $\gamma_b$ and $\eta$ were varied to find the most adequately fitted MWL spectrum for each cases of $t_{age}$. The resulting plot is given in the left panel of Figure \ref{fig6}.
The MWL spectrum gradually deviates away from the observed MWL SED, and consequently the fit worsens. Since the magnetic field increases due to the compression of the PWN during the reverberation phase, the chosen input of the magnetic fraction $\eta$ was decreased with increasing considered age, 
to control the fraction of spin down luminosity that goes on to power the magnetic field of the nebula.
Nevertheless, due to the compression of the nebula and high synchrotron burning, the power law index at high energy gradually softens with age, 
indicating efficient cooling of the injected high energy electrons to lower energies during reverberation. This in turn affects the spectrum of the resultant synchrotron and IC photons produced. This fact is apparent from the figure, as the calculated MWL spectrum is unable to explain the observed X-ray and the VHE-UHE data present in the SED.

We have also considered $t_{age}$ = 6500, 6800, 7200 and 7500 years and fitted the SED to study how the magnetic field and radius change. The corresponding plot is given in the right panel of Figure \ref{fig6}. The final PWN radius increases with increasing age, and also corresponds to the onset of the PWN contraction due to the reverberation in all cases. The PWN must be at the beginning stage of the contraction if the observed SED is to be explained. 
In all cases, the magnetic field is still low and close to, or lower than the Galactic average value. 

\subsection{$t_{age}$ as a free parameter}\label{subsec_4}

From the above discussion, it is apparent that the true age of the PWN can not be much greater than 7000 years, due to reverberation kicking in afterwards. As discussed in subsection \ref{subsec_1}, the true age can not also be lower than 4000 years. Thus, we have used \texttt{TIDEFIT}, leaving the true age parameter free.
Apart from $t_{age}$, the free parameters considered in this case are same as that discussed in subsection \ref{subsec_2}. 
The best-fit $t_{age}$ from the fitting of the MWL SED is 4880.3 (3871.4, 6060.4) years. This value is consistent with the range of the said parameter discussed above. The best fitted model spectrum is similar to that shown in Figure \ref{fig3}. Although the best-fit values of the free parameters, apart from t$_{age}$, are not exactly the same to those obtained assuming  $t_{age}$ = 7000 years, the 1$\sigma$ confidence intervals of these free parameters for the two cases, overlap with each other  (see subsection \ref{subsec_2} and Table \ref{tab1}). Both solutions are essentially the same within uncertainties. So we do not report the best fit results obtained in this case. We have also obtained a comparatively large PWN radius ($R_{PWN}$ = 8.26 pc) and low magnetic field ($B_{PWN}$ = 1.87 $\mu$G) associated with the PWN at the best fitted $t_{age}$, similar to the case discussed in subsection \ref{subsec_2} for t$_{age}$ = 7000 years, what we discuss next.

\section{Concluding Remarks: large radius/low magnetic field issue}\label{sec5}

In this paper, we have provided a detailed, time-dependent, one-zone model to explain the UHE gamma ray emission observed from the direction of LHAASO J2226+6057 using the emission from PWN associated with PSR J2229+6114.
We summarize the main points obtained in this work, and compare them with previous studies.

\begin{enumerate}
    \item The effects due to the variation of the true age and braking index were not explored in previous studies done for this source. 
    From our study, we found that a true age between 4000 and 7000 years is most suitable for the fitting of the SED of the source. {Considering $t_{age}$ as a free parameter during the fitting of the MWL SED also revealed a compatible result.} Our study found that no significant effects can be seen on the MWL spectrum if we take different braking indices. 
    
    \item We have used the \texttt{TIDEFIT} code to fit the observed MWL SED by computing the best $\chi^2$ fit model spectrum. 
    From the fit, we found that CMB is the target photon field responsible for IC cooling of the injected leptons from the PWN, whereas the effects of FIR/NIR radiation fields are  negligible in this case. 
    Additionally we have found that the PWN at current age must be extended ($R_{PWN} \sim$ 10 pc), as are the extended source regions observed by VERITAS ($\sim$ 14 pc (0.27$^{\circ}$)) \citep{acciari09} and LHAASO ($\sim$ 25.6 pc (0.49$^{\circ}$)) \citep{cao21}. {The obtained PWN radius is larger when compared to e.g., X-ray radii observed for the Boomerang PWN, which is normal in one-zone models.}

    \item We have taken the effect of reverberation into account in our modeling. It was found that the PWN is at the onset of compression due to the impact of reverse shock hitting the shell of the PWN. The effect of reverberation on the MWL spectrum was also explored by gradually increasing the true age of the PWN to the characteristic age. The MWL description worsened with increasing age. The true age can not be much larger than the considered age of 7000 years if a PWN is {responsible for the gamma ray emission detected}. Since reverberation does not provide  a better fit, its inclusion does not solve the {large radius/low magnetic field} issue.
    
    \item We have also found that a very low magnetic field ($\sim$ 2 $\mu$G) is needed to explain the MWL SED of the source, which is comparable with the Galactic average magnetic field value. A similarly low magnetic field (a few $\mu$G) was found in previous studies as the one needed to describe this source \citep{joshi22, yu22}. Moreover, a low magnetic field was found for other LHAASO detected PeVatron cadidates as well \citep{desarkar22, joshi22, crestan21, li21, burgess22}, which is an a priori obvious outcome of requesting a leptonically-generated 
    high energy emission. Further complications of the model, as we discussed, do not significantly alleviate this.
\end{enumerate}

{As discussed earlier, an uncertainty regarding the distance of the source remains. We have explored this uncertainty by considering two very different values $D = 800$ pc and $D = 7.5$ kpc. For $D = 7.5$ kpc, we have found the fit is comparatively worse compared to that discussed in subsection \ref{subsec_2}, with $\chi^2$/D.O.F. = 36.78/30 and systematic uncertainty of 0.41. 
On the other hand, the fit in the case of $D = 800$ pc is comparable to that discussed in subsection \ref{subsec_2}, with $\chi^2$/D.O.F. = 35.88/30 and systematic uncertainty of 0.32.  There is no clear evidence to overrule one distance over the other, so we only report in detail the case of $D = 3$ kpc to directly compare with the works of \cite{joshi22} and \cite{yu22}. It is to be noted that power law lepton injection spectrum is more favoured in case of $D = 800$ pc, rather than a broken power law spectrum. However, the issue of large radius/low magnetic field remains in this case as well.}

The large estimated PWN size appears to be a caveat of the PWN interpretation of the LHAASO source. It is likely that both the head and the tail region contribute to the total observed emission from the source. In light of such a complicated source morphology, a one-zone treatment of the source region considered in the model proves to be a simplistic take on the same. The large PWN radius obtained from the calculation could be a consequence of such simplistic assumption. Nevertheless, our model tends to be the most definitive PWN approach considered thus far to explain the MWL emission from the LHAASO source, complete with reverberation consideration and PWN true age estimation. It is to be noted that the PWN radius obtained by \cite{joshi22} ($\sim$ 3.1 pc) is different from that obtained in this work.  Although one zone model has been considered in both cases, the  differences in the formalism adopted to compute the radius evolution in presence of the background SNR may be the reason behind this. In any case, further investigation will be important to properly address the complexity of the source morphology.

The Galactic magnetic field (GMF) plays a important role in the cosmic ray propagation. The intensity and orientation of GMF are constrained by several methods such as Zeeman splitting observations \citep{crutcher99}, infrared, synchrotron and starlight polarization studies \citep{nishiyama10, jaffe10, heiles96}, and Faraday rotation measures \citep{han06, pshirkov11}. The GMF model typically has three components, namely the Disc ($B_{Disc}$), Halo ($B_{Halo}$) and Turbulent ($B_{Turb}$) contributions. 
Typical values for $B_{Disc}$ and $B_{Halo}$ lie within the range of 2-11 $\mu$G \citep{desarkar21, bernardo13}, although the value of the turbulent component depends on the halo height of the Galaxy \citep{bernardo13}, e.g. for a typical halo height of 8 kpc, the value of $B_{Turb}$ comes out to be $\approx$ 6 $\mu$G \citep{desarkar21}. Although the exact structure of the small scale GMF is not known yet, from the models given by \cite{pshirkov11} and \cite{farrar12}, and also from observed secondary-to-primary ratio (\cite{desarkar21}, and references therein), the large scale, average GMF can be estimated to be in the range 2-6 $\mu$G. The PWN magnetic field ($\sim$ 2 $\mu$G) would be marginally close to, or even lower than, any of the estimates of the average GMF value.

We see two possible ways out of this. On the one side, the local environment of the PWN could have been vacuated by the explosion of the supernova, and/or by earlier explosions, so that the local field in the vicinity is actually much lower than a few $\mu$G, allowing for a magnetic field contrast to appear between the PWN and its environment. 
On the other hand, there is still a chance that the approximate representation of reverberation we have adopted here is still misleading. We know that assuming the ejecta pressure as constant (as all PWNe models have so far done) is in fact an oversimplification \citep{Bandiera:2020, Bandiera:2022}. A better treatment of the ejecta pressure could plausibly change the reverberation results, and we shall explore this in the future.
Of course, it is also possible that the PWN explanation of the source is not realized at all, and/or that a hadronic component plays a dominant role.
For the moment, a conflicting low value for the PWN magnetic field (not solved by age, braking indices, or currently assumed behavior of the reverberation process) 
leaves the PWN origin of LHASSO J2226+6057, and other similar sources,
in search for further observational tests.

\begin{acknowledgements}
      This work has been supported by grant 
PID2021-124581OB-I00.
DFT acknowledges as well USTC and the Chinese Academy of Sciences Presidential International Fellowship Initiative 2021VMA0001. 
This work was also supported by the Spanish program Unidad de Excelencia María de
Maeztu CEX2020-001058-M. ADS acknowledges hospitality provided by the Institute of Space Sciences, where majority of this work was carried out. ADS also acknowledges RRI HPC2020 cluster for providing facilities to perform numerical computations reported in this work. ADS thanks Nayantara Gupta for helpful comments.
\end{acknowledgements}

%
\bibliographystyle{aa} 
\bibliography{aa.bib} 
%

\end{document}